\documentclass{aa}
\usepackage{graphicx}
\usepackage{txfonts}

\usepackage{diagbox}
\usepackage{rotating}
\usepackage{tikz}
\usepackage{multirow}
\usepackage{wasysym}
\usepackage{adjustbox}
\usepackage{afterpage}
\usepackage{placeins}
\usepackage{xcolor}
\usepackage{caption}
\usepackage[strict]{changepage}
\usepackage{lscape}
\usepackage{pdflscape}
\usepackage{soul}

\usepackage{hyperref}
\hypersetup{colorlinks=True,citecolor=blue}

\def\ms{\hbox{\,m\,s$^{-1}$} }         
\def\cms{\hbox{\,cm\,s$^{-1}$} }       
\def\m2s2{\hbox{\,m$^{2}$\,s$^{-2}$} } 
\def\kms{\hbox{\,km\,s$^{-1}$} }       
\def\vsini{\hbox{$v$\,sin\,$i$}}      

\def\logrhk{$\log$(R$^{\prime}_{HK}$) }

\begin{document}
   \title{Three Years of HARPS-N High-Resolution Spectroscopy and Precise Radial Velocity Data for the Sun}
   
   \titlerunning{Precision Radial Velocity of the Sun with HARPS-N}
   \authorrunning{X. Dumusque et al.}


   \author{X. Dumusque \inst{1}
          \and M. Cretignier\inst{1}
          \and D. Sosnowska \inst{1}
          \and N. Buchschacher \inst{1}
          \and C. Lovis \inst{1}
          \and D. F. Phillips \inst{2}
          \and F. Pepe \inst{1}
          \and F. Alesina \inst{1}
          \and L. A. Buchhave \inst{14}
          \and J. Burnier \inst{1}
          \and M. Cecconi \inst{7}
          \and H. M. Cegla \inst{1,12}
          \and R. Cloutier \inst{2}
          \and A. Collier Cameron \inst{3}
          \and R. Cosentino \inst{7}
          \and A. Ghedina \inst{7}
          \and M. Gonz\'alez \inst{7}
          \and R. D. Haywood \inst{11}
          \and D. W. Latham  \inst{2}
          \and M. Lodi \inst{7}
          \and M. L\'opez-Morales \inst{2}
          \and J. Maldonado \inst{13}
          \and L. Malavolta \inst{15}
          \and G. Micela \inst{13}
          \and E. Molinari \inst{6}
          \and A. Mortier \inst{4,5}
          \and H. P\'erez Ventura \inst{7}
          \and M. Pinamonti \inst{10}
          \and E. Poretti \inst{7,8}
          \and K. Rice \inst{9}
          \and L. Riverol \inst{7}
          \and C. Riverol \inst{7}
          \and J. San Juan \inst{7}
          \and D. S\'egransan \inst{1}
          \and A. Sozzetti \inst{10}
          \and S. J. Thompson \inst{4}
          \and S. Udry \inst{1}
          \and T. G. Wilson \inst{3}
          }

   \institute{Astronomy Department of the University of Geneva, 51 ch. des Maillettes, 1290 Versoix, Switzerland\\
              \email{xavier.dumusque@unige.ch}
         \and Center for Astrophysics | Harvard \& Smithsonian, 60 Garden Street, Cambridge, MA 02138, USA
         \and SUPA School of Physics and Astronomy, University of St Andrews, North Haugh, St Andrews KY16 9SS, UK
         \and Astrophysics Group, Cavendish Laboratory, University of Cambridge, J.J. Thomson Avenue, Cambridge CB3 0HE, UK
         \and Kavli Institute for Cosmology, University of Cambridge, Madingley Road, Cambridge CB3 0HA, UK
         \and INAF - Osservatorio Astronomico di Cagliari, via della Scienza 5, 09047, Selargius, Italy
         \and Fundaci\'on Galileo Galilei-INAF, Rambla Jos\'e Ana Fernandez P\'erez 7, E-38712 Bre\~na Baja, TF, Spain
         \and INAF - Osservatorio Astronomico di Brera, via E. Bianchi 46, I-23807 Merate (LC), Italy
         \and SUPA, Institute for Astronomy, University of Edinburgh, Blackford Hill, Edinburgh, EH9 3HJ, Scotland, UK
         \and INAF—Osservatorio Astrofisico di Torino, Strada Osservatorio 20, Pino Torinese (To) I-10025, Italy
         \and Astrophysics Group, University of Exeter, Exeter EX4 2QL, UK
         \and Department of Physics, University of Warwick, Coventry, CV4 7AL, UK
         \and INAF – Osservatorio Astronomico di Palermo, Piazza del Parlamento 1, 90134 Palermo, Italy
         \and DTU Space, National Space Institute, Technical University of Denmark, Elektrovej 328, DK-2800 Kgs. Lyngby, Denmark
         \and Department of Physics and Astronomy, Universit\'a degli Studi di Padova, Vicolo dell’Osservatorio 3, I-35122, Padova, Italy 
             }

   \date{Received XXX ; accepted XXX}

 
  \abstract
   {The solar telescope connected to HARPS-N has been observing the Sun since the summer of 2015. Such high-cadence, long-baseline data set is crucial for understanding spurious radial-velocity signals induced by our Sun and by the instrument. On the instrumental side, this data set allowed us to detect sub-\ms\,systematics that needed to be corrected for.}
  {The goal of this manuscript is to i) present a new data reduction software for HARPS-N, ii) demonstrate the improvement brought by this new software on the first three years of the HARPS-N solar data set, and iii) release all the obtained solar products, from extracted spectra to precise radial velocities.}
   {To correct for the instrumental systematics observed in the data reduced with the current version of the HARPS-N data reduction software (DRS version 3.7), we adapted the newly available ESPRESSO DRS (version 2.2.3) to HARPS-N and developed new optimized recipes for the spectrograph. We then compared the first three years of HARPS-N solar data reduced with the current and new DRS.}
   {The most significant improvement brought by the new DRS is a strong decrease in the day-to-day radial-velocity scatter, from 1.27 to 1.07\ms; this is thanks to a more robust method to derive wavelength solutions, but also to the use of calibrations closer in time. The newly derived solar radial-velocities are also better correlated with the chromospheric activity level of the Sun on the long-term, with a Pearson correlation coefficient of 0.93 compared to 0.77 before, which is expected from our understanding of stellar signals. Finally, we also discuss how HARPS-N spectral ghosts contaminate the measurement of the calcium activity index, and present an efficient technique to derive an index free of instrumental systematics.}
  { This paper presents a new data reduction software for HARPS-N, and demonstrates its improvements, mainly in term of radial-velocity precision, when applied to the first three years of the HARPS-N solar data set. Those newly reduced solar data, representing an unprecedented time-series of 34550 high-resolution spectra and precise radial velocities, are released alongside with this paper. Those data are crucial to understand further stellar activity signals in solar-type stars and develop the mitigating techniques that will allow us to detect other Earths.}


   \maketitle

\section{Introduction}

The radial-velocity (RV) method is an efficient technique to estimate the mass of exoplanets.
If the planets are also transiting, their radius and exact mass can be obtained, thus revealing the exoplanet's density, leading
to planetary interior and atmospheric characterisation. The RV technique is also currently the only method for which available instruments reach the sensitivity needed to detect low-mass, 
non-transiting exoplanets orbiting stars within 40 parsecs from the Sun. Those planets will be prime targets 
for future direct imaging mission \citep[e.g.][]{nas2018} such as LUVOIR (\url{https://asd.gsfc.nasa.gov/luvoir/resources/}) and HabEx (\url{https://www.jpl.nasa.gov/habex/documents/}).

However, when looking for Earth-mass planets orbiting in the habitable zone of solar-type stars, 
the RV technique is limited by spurious radial-velocity signals induced by the host star at the \ms\,level \citep[e.g.][]{Fischer-2016, Plavchan:2018aa, eprvwg}.
It is therefore crucial to obtain a better understanding of stellar signals, in order to mitigate them and thus reveal tiny planetary signatures with amplitudes as small as 10\cms.

Due to stellar signal timescales ranging from minutes for stellar oscillations to years for magnetic cycles, standard 
radial-velocity observations obtained either from transit follow-up observations or blind search surveys very often lack the sampling and time baselines necessary 
to characterise the different types of stellar signals. The lack of good data makes it extremely difficult to understand stellar signals in enough detail, and therefore to
find efficient techniques to mitigate them. The ideal solution to make progress 
is to continuously observe a target which astrophysical parameters and activity behaviour as a function of time are perfectly known, with the same instrument employed for exoplanet characterization. This reflection led to the development of the HARPS-N low-cost solar telescope \citep[LCST,][]{Dumusque-2015b}. Since then, several solar feeds have been 
or will be installed at different sub-\ms RV precision facilities (HARPS, EXPRES, NEID, MAROON-X, private communication with the different instrument teams). We note that a solar feed is also present on the 
PEPSI spectrograph \citep[][]{Strassmeier-2015}, and on the NIR spectrograph GIANOB \citep[][]{Claudi:2018aa}, however, the obtained data are not dedicated to extreme RV precision.

Since July 2015, HARPS-N has been observing the Sun for several hours nearly every day, with a cadence of 5 minutes, using its solar 
feed \citep[][]{Phillips:2016aa}. Since then, several papers have analysed the obtained data and investigated solar radial-velocity signals and related
activity-index indicators \citep[][]{Collier-Cameron2019aa, Milbourne:2019aa, Maldonado:2019aa, Miklos:2020aa,Langellier:2020aa}

The excellent quality of the solar RVs published in \citet{Collier-Cameron2019aa}, with instrumental systematics below the \ms\,level, and the exceptional cadence of these data over several years, revealed instrumental systematics that 
needed to be understood and corrected for. The first step in this direction was to adapt the publicly available ESPRESSO Data Reduction 
Software \citep[DRS,][]{Pepe:2020aa} to HARPS-N so that the solar data could be reduced with an up-to-date DRS. In this paper, we will refer to the current HARPS-N 
DRS (version 3.7) as the old DRS and the ESPRESSO pipeline(version 2.2.3) adapted to HARPS-N as the new HARPS-N DRS. Beyond using the ESPRESSO DRS to extract RV information from raw spectra, we also developed new recipes to specifically account for
HARPS-N systematics. We implemented a new procedure to derive wavelength solutions, optimised the selection of thorium lines
to improve RV stability on the long-term, used master flat-fields to prevent being photon-noise limited by them when analysing solar data, used the calibrations the closest in time, and corrected for ghosts contamination in the calcium activity index time-series.

The article is structured as follow. In Sect.~\ref{RV_improvements} we discuss the optimisation of the ESPRESSO DRS to HARPS-N and the new recipes implemented to solve for known systematics. In Sect.~\ref{RV_data_set}, we discuss the HARPS-N solar data set used in this paper, we compare the solar RVs reduced with the old and new DRS, and we describe the data that are released alongside with this paper. Finally, we discuss the results found and conclude in Sect.~\ref{Conclusion}.

\section{Deriving precise HARPS-N radial-velocity measurements} \label{RV_improvements}

\subsection{The ESPRESSO DRS} \label{espresso_drs}

To improve the precision of the HARPS-N RVs, we first adapted the publicly available ESPRESSO DRS
to work with HARPS-N. Although many of the data reduction steps and algorithms implemented in the ESPRESSO
DRS find their origins in the HARPS-N DRS, several
steps were significantly re-designed and improved with respect
to HARPS-N to meet the more challenging requirements of
ESPRESSO. A comparison of the old HARPS-N and new ESPRESSO DRS is beyond the scope of this paper, and interested readers are encouraged to read \citet{Pepe:2020aa} for a general description and in the near future, the paper describing specifically the ESPRESSO DRS. 

\subsection{Improved wavelength solution} \label{sect:new_wave_sol}

In the old HARPS-N DRS, a wavelength solution for the science fibre is performed every day by illuminating the corresponding fibre with a thorium-argon lamp.
A list of thorium lines, 3692 in total, with their physical position on the detector and their laboratory wavelength is used to detect and measure the centroid of each thorium line in the extracted calibration spectrum. Readers interested in how those lines have been selected should have a look at \citet{Lovis-2007b}. Using the line centroids and corresponding laboratory wavelengths, a wavelength for each pixel is derived by interpolation. For each order individually, this interpolation is performed using a third order polynomial. The old HARPS-N DRS, like the new one, extracts 69 orders from the raw images; thus, wavelength solutions are derived using a total of $4\times69=276$ free parameters. 

We note that to ease the comparison with the DRS data products, we will use throughout the manuscript the order numbering of the DRS. Order 0 thus corresponds to the first row in the echelle-order spectra derived from the DRS, order 68 to the last one. In terms of physical orders, order 0 of the DRS correspond to physical order 158 (average wavelength 6873 \AA) and order 68 to physical order 90 (average wavelength 6872 \AA).

\begin{figure*}[h]
	\includegraphics[width=18.5cm]{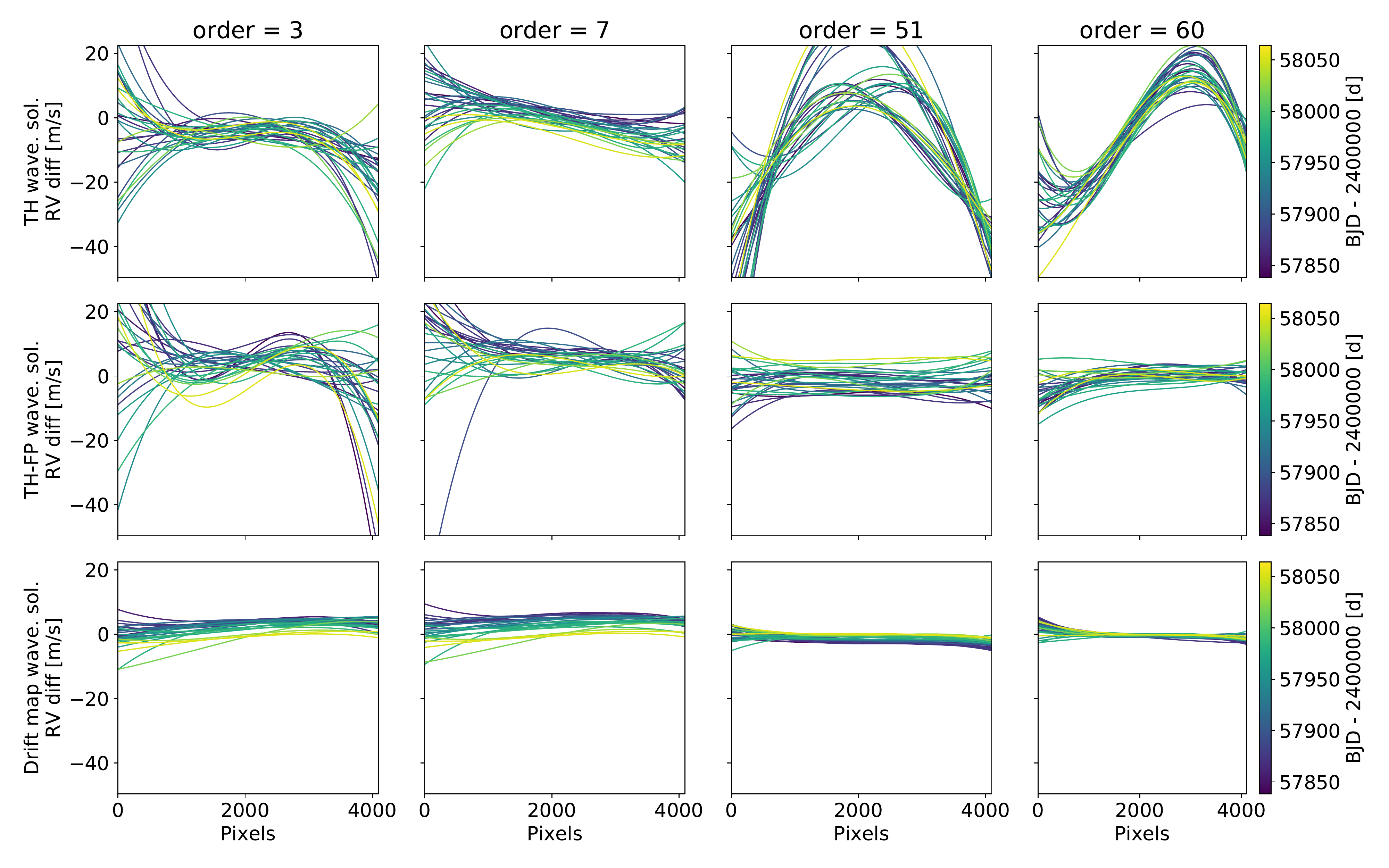}
	\caption{RV difference between wavelength solutions taken on different days when derived using different algorithms. The RV difference is obtained by comparing all the wavelength solutions with the wavelength solution of reference used in the drift map method, obtained on 2017-11-11. Each column corresponds to a different echelle order, highlighting the behavior of the wavelength solutions from the blue (order 3) to the red (order 60).   \emph{Top:} RV difference when using only the thorium lines (TH). This is what is done in the old HARPS-N DRS. \emph{Middle:} RV difference when using thorium lines and the Fabry-P\'erot spectrum (TH-FP). \emph{Bottom:} RV difference when using the drift map method presented in this paper, which means using a reference wavelength solution, and shifting it in velocity after measuring the drift of the thorium lines.}
	\label{fig:wave_sol_diff}
\end{figure*}

The flux throughout an echelle order is not constant due to the blaze response, and is very low on both sides of the order. Therefore, thorium lines in the center of an order are much more constraining for the S/N-weighted polynomial fit used to derive the wavelength solution than thorium lines on both sides of the order. This gives rise to instability in the fitted polynomials and therefore in the wavelength solution from calibration to calibration, which are generally performed every day. This problem was already discussed in several papers \citep[][]{Dumusque:2018aa, Cersullo-2018, Bauer:2015aa} and we show in the top row of Fig.~\ref{fig:wave_sol_diff} an example of how the wavelength solution for a few orders can differ by tens of meters-per-second from one calibration to the next. We show here several wavelength solutions spread over 200 days, and we observe that these strong variations can be seen from one calibration to the next, even when taken only a few minutes apart. This instability in wavelength solutions translates into day-to-day offsets in RV measurements with an estimated root-mean-square (rms) of about 0.4\ms\,\citep[][]{Dumusque:2018aa}.
%

To solve for this instability issue in the wavelength solution, the ESPRESSO DRS uses a combination of a thorium calibration on the science fibre, followed by a Fabry-P\'erot calibration on the same fibre to stabilise the wavelength solution. The thorium lines give the absolute wavelength scale, while the Fabry-P\'erot's very rich spectrum, with each emission peak being separated by the same distance due to the interferometric nature of the emission, allows us to better constrain the polynomial fit. Readers interested in how to implement such a technique are directed to \citet{Cersullo-2018} and \citet{Bauer:2015aa}. However, for HARPS-N, such set of calibrations is not present during the entire lifetime of the instrument. As we wanted to adopt a single wavelength solution derivation method for the spectrograph, we decided to implement another strategy, that we call the drift map method. The idea is to use a reference wavelength solution, obtained at a time when consecutive thorium and Fabry-P\'erot calibrations on the science fibre are available, and then measure the drift of the thorium lines between each thorium calibration and the reference thorium calibration. A two-dimensional polynomial fit is then performed on the measured drift of the thorium lines and a wavelength solution is obtained for each thorium calibration by shifting accordingly the reference wavelength solution. A detailed description on how to derive a wavelength solution using the drift map method can be found in Appendix~\ref{app:drift_map_method}. 

To test how the drift map method compares to the thorium-only (old HARPS-N DRS) and combined thorium--Fabry-P\'erot techniques when deriving wavelength solutions, we selected 29 thorium calibrations on the science fibre, followed by Fabry-P\'erot calibrations on the same fibre (from 2017-03-25 to 2017-11-06). We then derived the wavelength solutions for those calibrations using the thorium--Fabry-P\'erot technique and the drift map method, and compared the obtained products to the wavelength solutions derived using the old HARPS-N DRS. In Fig.~\ref{fig:wave_sol_diff}, we show the RV drift, in meters-per-second, between each of these wavelength solutions and the reference wavelength solution used in the drift-map method, obtained on 2017-11-11. From top to bottom, we show the RV drift for the wavelength solution derived using the old HARPS-N DRS, the thorium--Fabry-P\'erot combined solution, and the drift map method, respectively. Each column in the figure corresponds to different orders from the blue to the red, including echelle orders 3, 7, 51 and 60. We note that during those 8 months, the instrument drifted by about $+15$\ms, which induces an increasing RV offset when going from the 2017-03-25 to the 2017-11-06 wavelength solutions (purple to yellow colored lines in Fig.~\ref{fig:wave_sol_diff}). However, we measured this instrumental drift on the drift map wavelength solutions and corrected it for all wavelength solutions. To obtain a mean drift value for each drift map wavelength solution, we performed a weighted average taking into account the blaze response of the instrument, and the change of S/N across the echelle orders.

Looking at Fig.~\ref{fig:wave_sol_diff}, it is clear that the thorium only wavelength solutions show more variations from one wavelength solution to the next than the thorium-Fabry-P\'erot and drift map wavelength solutions for orders 51 and 60. For blue orders 3 and 7, the thorium-Fabry-P\'erot solutions shows similar variations as the thorium only solutions, and only the drift map method gives more stable results. This comes from the fact that the flux on the HARPS-N  Fabry-P\'erot is rather low in the ten first orders of the spectrograph. Fabry-P\'erot emission peaks on the borders of these blue orders are difficult to detect due to low S/N, and therefore we needed to add the information provided by the thorium lines to improve the wavelength solutions in the blue. Without adding the information from the thorium lines, the thorium-Fabry-P\'erot wavelength solutions in the blue part of HARPS-N are even worse than the thorium only solution. We note that the HARPS-N Fabry-P\'erot was upgraded after 2017-11-11 and the Fabry-P\'erot spectrum now has much more flux in the blue part of the spectrograph, as is also the case for ESPRESSO. For all the orders, from the extreme blue to the extreme red, we see that the wavelength solutions derived with the drift map method are the most stable ones. We note that although we see that the drift map wavelength solutions present the smallest variations, this is biased by the fact that rather than fitting $276$ free parameters to derive a wavelength solution, our drift map model only includes 15 parameters (see Appendix~\ref{app:drift_map_method}). In the absence of an absolute wavelength solution derived using a laser frequency comb, it is impossible to say which
of the thorium-Fabry-P\'erot or drift map wavelength solutions are the more precise for echelle orders greater than $\sim$10. Since the drift map method gives wavelength solutions much more stable than the old DRS for all the echelle orders, we decided to implement this technique in the new HARPS-N DRS.

\begin{figure}
        \center
	\includegraphics[width=9cm]{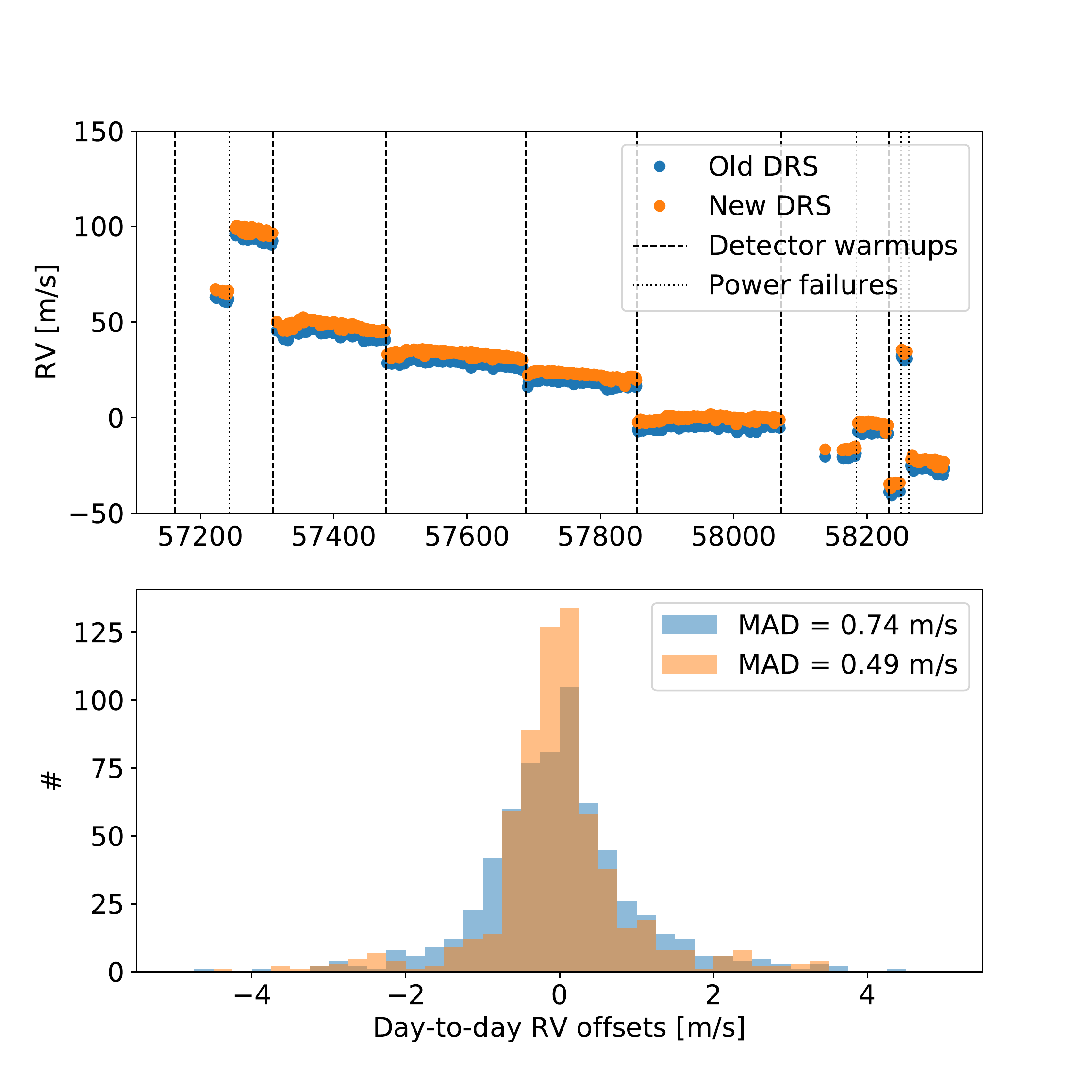}
	\caption{RV stability of the wavelength solutions derived with the old (blue) and new (orange) DRS. \emph{Top: }RV drift of HARPS-N over-time, measured by computing the RV drift between all wavelength solutions and the wavelength solution of reference (set for day 2017-11-11, see text). We see a general drift over time, interrupted by significant jumps at each warm-up of the detector and power failures (dashed and dotted vertical lines). \emph{Bottom: }Distribution of the RV offsets measured between consecutive wavelength solutions separated by less than 1.5 days, which removes the large offsets observed at each instrument intervention. The MAD of these day-to-day RV offsets is 0.74 and 0.49\ms for the old and new DRS, respectively.}
	\label{fig:inst_drift}
\end{figure}

Wavelength solutions set the RV zero-point of the spectrograph each time a new calibration is obtained, on average every day. Thus, their precision impact directly the stability of the measured RVs from one day to the next. To measure the gain in RV stability brought by the new drift map wavelength solutions compared to what was done in the old HARPS-N DRS, we measured the mean RV drift of all wavelength solutions relative to the reference wavelength solution used (set for day 2017-11-11, see above). Like in the analysis performed above, the mean RV drift is measured using a weighted average taking into account the blaze response of the instrument and the varying S/N across echelle orders. In the top panel of Fig.~\ref{fig:inst_drift}, we show the result for the old and new wavelength solutions. We clearly see that the old and new DRS give a similar HARPS-N drift over three years, of a hundred meters-per-second. We also observe sudden jumps at each instrument interventions, either corresponding to a scheduled detector warm-up or to a power failure. In both cases, the observed offset is likely because after warming up and cooling down the detector, it does not come back exactly at the same previous position (see Sect.~\ref{correcting_calcium} and Fig.~\ref{fig:ghost_contam} for the rational behind periodically warming-up the detector). 

Over the long-term, several effects can influence the RV stability of the spectrograph, including the ageing of the thorium-argon lamp and the change of the focus of the instrument due to ageing of the optics. In the absence of an absolute calibrator as a laser frequency comb, we can only measure the precision in wavelength solution over short-timescales, considering that the instrument did not drift. To do so, we measured the RV offset between consecutive wavelength solutions that are separated by less than 1.5 days. This selection in time-separation allows to reject the large RV offsets measured between wavelength solutions taken before and after an instrument intervention. In the bottom panel of Fig.~\ref{fig:inst_drift}, we show the distribution of these measured day-to-day RV offsets. As we can see, the new DRS gives a distribution that is more peaked around zero, and that presents less outliers. The median absolute deviation (MAD\footnote{We note that all the values given for the MAD in this paper have been obtained from the formulae $\mathrm{median}\left[abs(y-\mathrm{median(y)})\right]/0.6744$, where the division by the constant 0.6744 allows to give a value comparable to a rms.}) of the two distribution is 0.74 and 0.49\ms for the old and new wavelength solutions, respectively. Consequently, by using the drift map method to derive wavelength solutions, we improve the stability of the RV measurements by a non-negligible $\sqrt{0.74^2-0.49^2}=0.55$\ms. We note that in \citet[][see appendix C]{Dumusque:2018aa}, a similar comparison using HARPS stellar data showed that the drift map wavelength solutions improved the RV precision by $0.4$\ms, a value slightly lower than what is found here for HARPS-N.
%

\subsection{New set of thorium lines for wavelength solutions}\label{sect:th_line_selection}

\begin{figure*}
        \center
	\includegraphics[width=9cm]{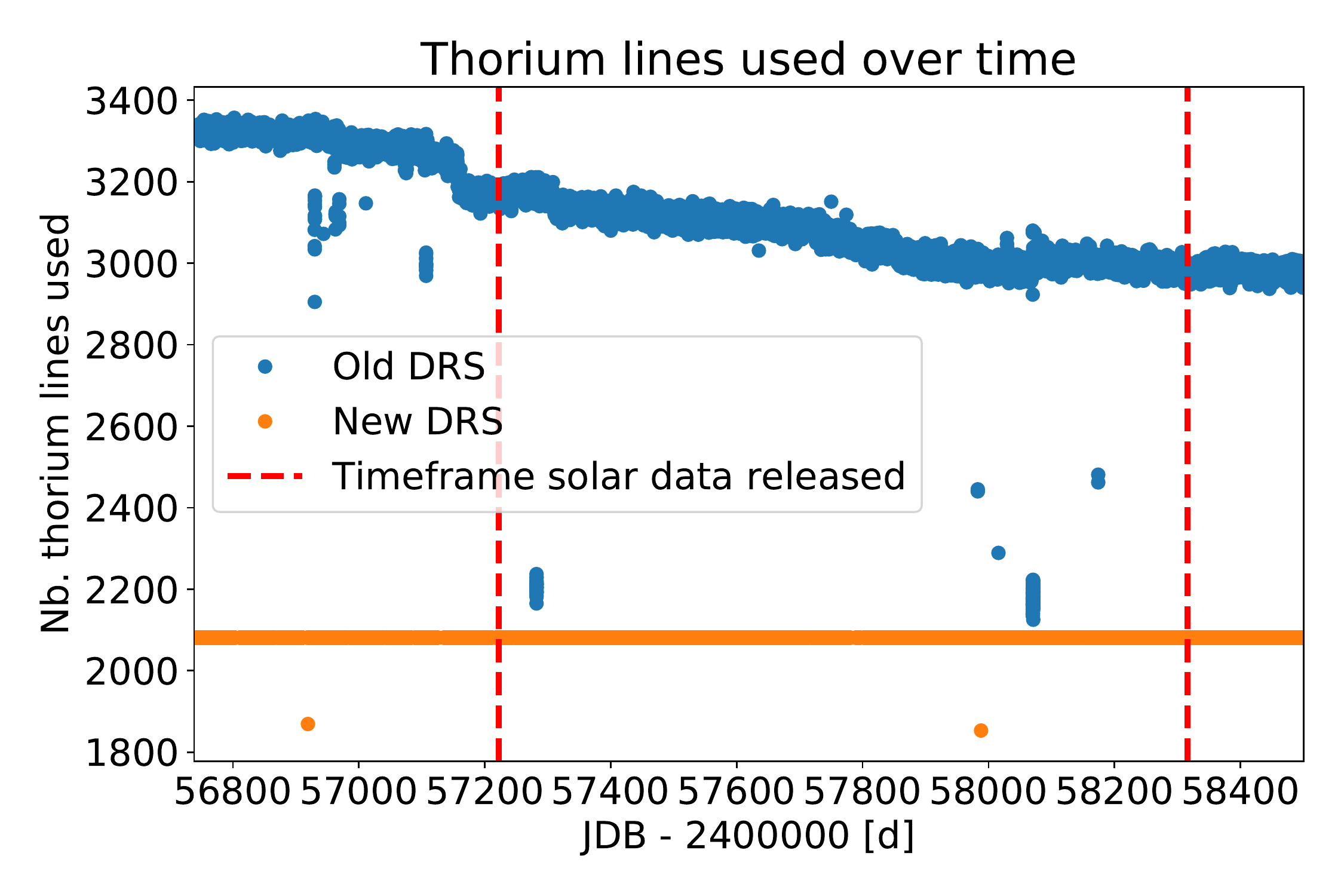}
	\includegraphics[width=9cm]{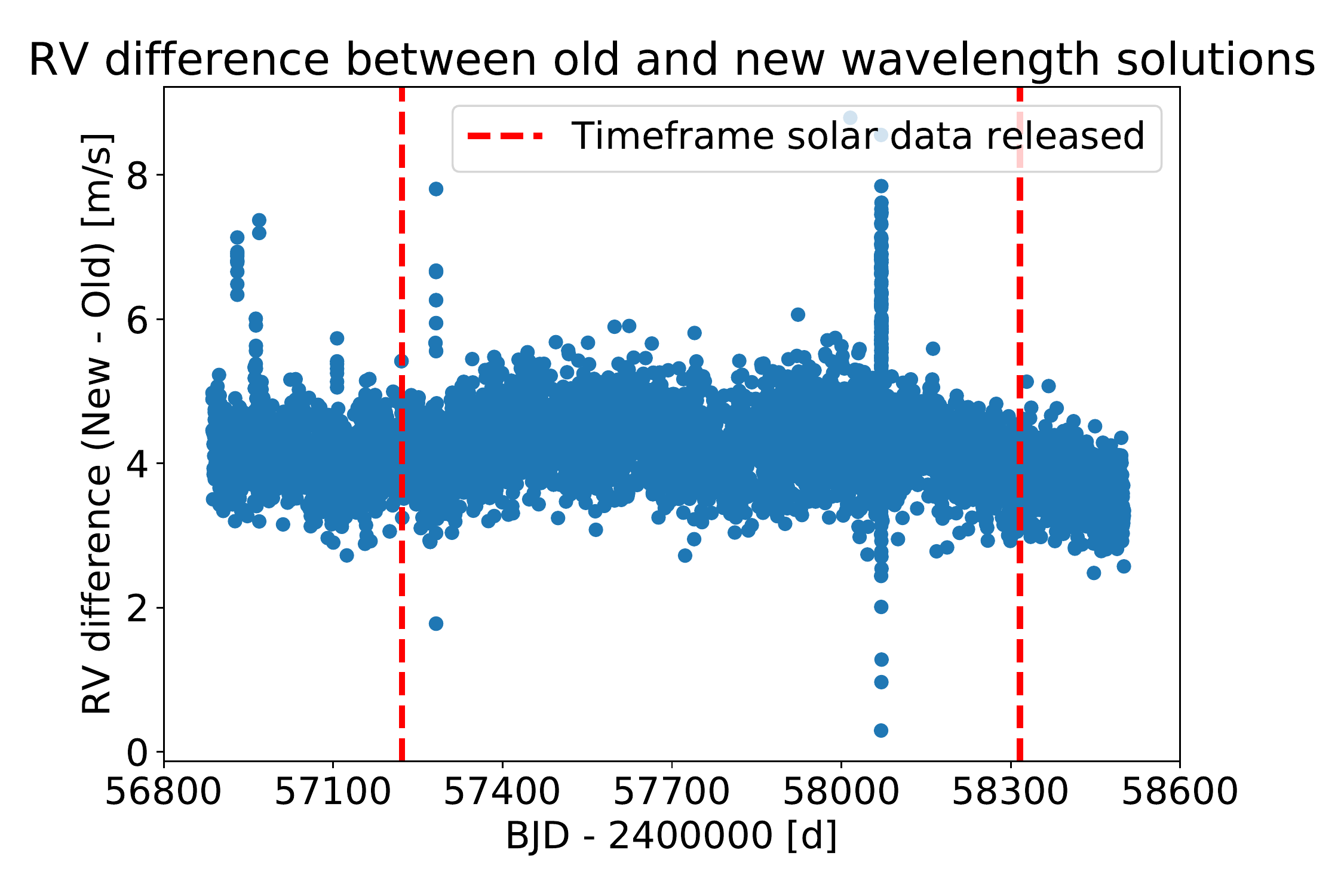}
	\caption{\emph{Left: }Number of thorium lines used over time to perform the wavelength solution in the old and new HARPS-N DRS. We show here the results for all thorium-thorium and thorium-Fabry-P\'erot calibrations. We see that there is a significant variation in the number of used lines in the old HARPS-N DRS, 16\%, while in the new HARPS-N DRS, we nearly always use 2041 thorium lines. \emph{Right: }RV difference between the wavelength solutions obtained with the old HARPS-N DRS and the new one, which uses the drift map method with the new set of thorium lines.} 
	\label{fig:th_lines_used_and_wavesol_compa}
\end{figure*}

After fitting for the position of the thorium lines, the old and new DRS check if lines are saturated and compare the fitted positions with reference positions. Lines that show a position significantly different from the reference are rejected, to prevent biasing the final wavelength solution. 
With ageing of the thorium lamp over time, more and more lines are rejected due to lines that start to saturate and lines for which blends change in intensity or in position\footnote{If a blend of a thorium line is from argon origin, this blend will be more sensitive than the thorium line to pressure changes in the lamp, and therefore will drift relative to it.}, which will induce a line-profile asymmetry and therefore a spurious RV drift. In the left panel of Fig.~\ref{fig:th_lines_used_and_wavesol_compa}, we can see that out of a total of 3332 thorium lines used by the old HARPS-N DRS on 2014-08-16 (BJD$=2456886$), only 2794 (16\% less) were still used 5.5 years later\footnote{We note that the same thorium-argon lamp was used during all this time.}, on 2019-01-16 (BJD$=2458500$). In addition, we will see in Appendix~\ref{app:th_line_selection} that the old HARPS-N DRS was using thorium lines for which the measured RV is strongly correlated with the flux variation of the thorium-argon lamp, thus pointing towards spurious RV shifts. Finally, the old HARPS-N DRS was fitting a combination of several Gaussian profiles on main thorium lines to account for blends, as described in \citet{Lovis-2007b}, while the new HARPS-N DRS fits only a single Gaussian profile for each thorium spectral line. The fact that the number of thorium lines used changes over time and that the new DRS does not fit anymore for blends can have a significant impact on the long-term RV stability. We therefore decided to perform a tailored selection of thorium lines for the new HARPS-N DRS. The description of this selection is rather technical and can be found in Appendix~\ref{app:th_line_selection}.
%

After selecting carefully thorium lines, we are left with 2041 lines that can be used to derive wavelength solutions based on the drift map method. Those thorium lines are nearly always used throughout the lifetime of HARPS-N (99\% of the time, see Appendix~\ref{app:th_line_selection}). We note that the entire study performed on the drift map wavelength solutions in Sect.~\ref{sect:new_wave_sol} uses this new set of thorium lines. 

In the right panel of Fig.~\ref{fig:th_lines_used_and_wavesol_compa}, we plotted the RV difference between the wavelength solutions obtained with the old HARPS-N DRS and the new one, which uses the drift map method with the new set of thorium lines. Overall, we see an offset of 4\ms between the wavelengths solutions, which will impact the accuracy\footnote{We note however that the RV accuracy is impacted by the uncertainty of much larger astrophysical effects such as the convective blueshift or the gravitational redshift, both of dozens of meters-per-second for solar-type stars.} of the derived RVs, but not their precision, which is generally what we want. At the second order level, we see that the RV difference between the wavelength solutions is slightly correlated (Pearson correlation coefficient of 0.39) with the thorium flux ratio (see top panel of Fig.~\ref{fig:thorium_analysis}), which measure the total flux emitted by the thorium-argon lamp at a given time relative to a reference time. As the thorium flux ratio significantly increases after BJD=2458070, the difference in RV is also affected and present a slope of $\sim$0.7$m\,s^{-1}\,yr^{-1}$. This tells us that the old or the new DRS seems more sensitive to flux variations of the thorium-argon lamp, however, without telling us which one it is. This can only be done by looking at the long-term monitoring of standard stars, what we will do for the Sun in Sect.~\ref{RV_comparison}.

\subsection{Master flat-fielding}  \label{sect:master_flat_field}

In standard HARPS-N calibrations, five consecutive frames with a tungsten lamp illuminating both science and reference fibres are used to derive the flat field for each echelle order and the corresponding blaze.
For most of the night-time observations, even when a few spectra are combined together, those flat-fields have a significantly high S/N to prevent adding noise when spectra are corrected from flat-field during the extraction process. For the Sun, however, as the median number of observations per day is 56, the S/N of the extracted flat field is a limitation when combining spectra reduced with the same calibrations (i.e. spectra taken on the same day). This is, for example, seen in an analysis performed in \citet[][see Fig.~8 and related discussion]{Cretignier:2020aa}. We therefore decided to create master flat fields that include more than five raw frames, to increase the S/N and thus be able to work with daily averages of spectra without being significantly affected by the noise present in the flat field products. This implies using frames from different days, which assumes that the instrument flat field does not change significantly over short periods of time. 

\begin{figure*}
        \center
	\includegraphics[width=16cm]{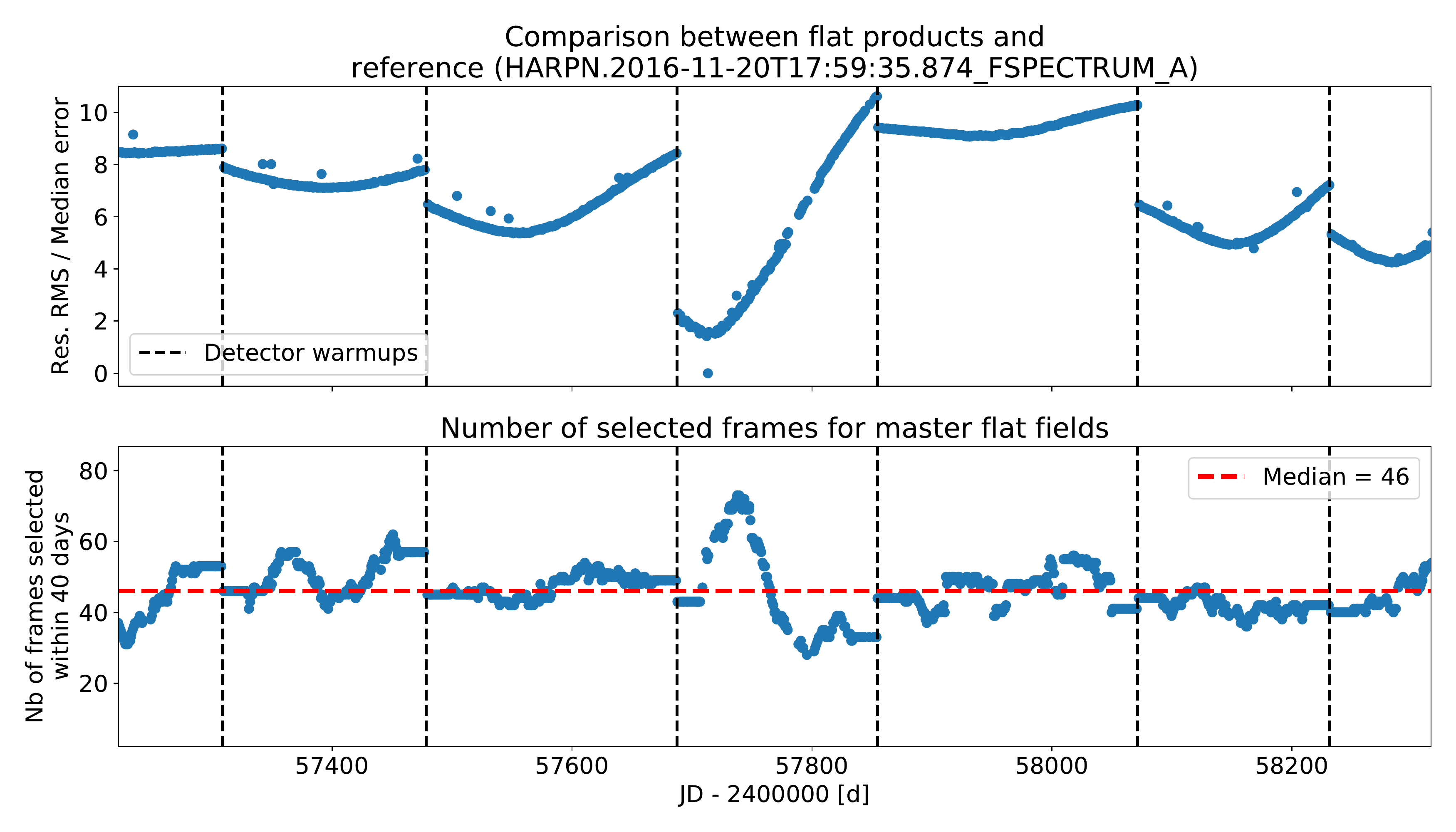}
	\caption{\emph{Top: }Variation of the flat fields over time. We show here the ratio between the rms and the median error of the residuals obtained by dividing all the flats by one of reference (HARPN.2016-11-20T17:59:35.874\_FSPECTRUM\_A.fits). Sudden jumps can be explained by periodic detector warm-ups (dasehd vertical lines). \emph{Bottom: }Number of raw frames used to build the master flat fields. This number has to be compared to 5, which is the number of raw frames included in the old HARPS-N flat field products, and 56, which is the median number of solar observation per day.}
	\label{fig:flat_stability}
\end{figure*}

To test the stability of flat fields over time, we divided all the extracted flat fields by a reference one, in our case HARPN.2016-11-20T17:59:35.874\_FSPECTRUM\_A.fits, and analysed if the residuals were following a Gaussian statistic. To do so, we measured the ratio between the rms and the median error of the residuals for each order. We note that the errors were propagated from the FLAT calibrations. Looking at the top panel of Fig.~\ref{fig:flat_stability}, where we show this ratio, we see that this observable varies slowly over time, with sudden jumps that are due to periodic warm-ups of the detector, highlighted by vertical dashed lines (see Sect.~\ref{correcting_calcium} for more details about these warm-ups). Between warm-ups, the variation of the flat field over time is small and it is therefore possible to increase the S/N by accumulating frames over several days. However, we do not want to combine frames on both sides of a detector warm-up. Therefore, we used the following strategy to select frames for a given master flat field. To build a master flat field for date YYYY-MM-DD, we looked at all the good raw frames within a time window of $\pm20$\,days. If there is a detector warm-up closer than 20 days, for example at $+10$\,days, we shift the window accordingly to still have a full window of 40 days. We thus consider the raw frames between dates YYYY-MM-DD$-30$\,days and YYYY-MM-DD$+10$ days. In this way, we will always keep a high number of frames in master flat fields, with the drawback that master flat fields closer than 20 days from a warm-up will all be identical. This is however not a problem due to the slow variation of the instrument flat field.

The raw frames used for flat fielding on HARPS-N have both fibres illuminated by the tungsten lamp. Therefore, the flat from the science fibre will be contaminated by ghosts from the reference fibre, which are due to secondary reflection of the echelle orders, which will not necessarily be the case of a science observation with no flux, the sky, the thorium-argon or the Fabry-P\'erot spectrum on the reference fibre. It is thus better to use calibrations where the tungsten lamp is only illuminating one fibre at a time, which prevents of having the ghosts from the other fiber (technique adopted on ESPRESSO). On HARPS-N, such calibrations exist and are generally used for order localisation, but nothing prevents us from using those raw frames to derive the flat field. The only problem is that there is only two such frames per calibration set (two for the science and two for the reference fibres), compared to the five frames available for standard flat-fielding. However, this can be compensated for by using frames from several different days. In the lower panel of Fig.~\ref{fig:flat_stability}, we show the number of localisation raw frames included in each master flat field. As we can see, the median number of frames is 46, with highest and lowest values at 28 and 73, which is similar to the median number of solar frames taken every day (56). Thus, when using the master flat fields described above to reduce solar data, the noise coming from flat fielding will not significantly affect daily averages of spectra. We note that all the data released with this paper have been reduced using those master flat fields.
%

\subsection{Use of afternoon calibrations} \label{sect:afternoon_calib}

In the old HARPS-N DRS, science data are reduced using the closest good quality calibrations back in time. This makes sense for night-time observations, as calibrations are performed in the late afternoon.
However, for daily observation of the Sun, this implies that the calibrations used are from the afternoon of the previous day, therefore at minimum more than 12 hours apart. This is not critical for order localisation, flat-fielding (see Sec.~\ref{sect:master_flat_field}), background and contamination corrections, however, it can be problematic for wavelength solutions, which set the RV zero point of the instrument every day and therefore directly impact RV stability. For each solar observation, with the Fabry-P\'erot spectrum on the reference fibre, the wavelength solution is derived by taking the wavelength solution calibration (thorium-argon lamp illuminating the science fibre and Fabry-P\'erot spectrum on the reference fibre) and shifting it by the Fabry-P\'erot drift observed on the reference fibre. However, this assumes that the Fabry-P\'erot spectrum is stable over time, which is not necessary the case. The first HARPS-N Fabry-P\'erot, that was in operation from the beginning of HARPS-N in 2012 to 2017-11-11, was showing internal drifts of a few dozen of \cms\,per day on average, with occasional sudden jumps that could be as high as 1 to 2\ms. This led to its replacement in 2017-11-11 by a more stable system free of instabilities and showing reduced daily drifts. We therefore expect HARPS-N data taken after 2017-11-11 in simultaneous Fabry-P\'erot mode to present less instrumental systematics.

The residual internal drift of the Fabry-P\'erot implies that, in the used calibration scheme, the wavelength solution is potentially affected by a systematic error that increases with the temporal distance between the calibration and scientific exposures. While in the old DRS RVs, the data were reduced using the calibration frames of the preceding afternoon, we reduced all the solar data presented in this paper by using the closest calibrations in the future, which are the ones obtained after the solar exposures on the late afternoon of the same day. This allows us to better mitigate the systematics induced by the internal drift of the Fabry-P\'erot. We did not want to use the closest calibration in time, as morning calibrations, sometimes performed, would induce a jump in the RVs observed within a day, as the wavelength solution calibration used would change from the morning one to the afternoon one around mid-day.

\subsection{Correcting for the calcium activity index contamination} \label{correcting_calcium}

The Mount-Wilson S-index, which is a measure of the chromospheric emission in the core of the Ca II H and K spectral lines, is one of the best activity proxies for solar-type stars.
On the Sun, it correlates extremely well with long-term changes in total solar irradiation induced by the presence of magnetically active regions such as faculae and sunspots \citep[e.g.][Fig.2]{Lockwood-2007}. For other stars, this index is used to track the rotational
period on the short term \citep[e.g.][]{Boisse-2009} and magnetic cycles on the long term \citep[e.g.][]{Wilson-1978, Baliunas-1995, Lovis-2011b, Dumusque-2012}. To be able to compare the activity of different stars, it is essential to correct the Mount-Wilson S-index from the photospheric flux contaminating the chromospheric emission in the core of the Ca II H and K spectral lines, which is performed when computing the \logrhk\,activity index \citep[][]{Noyes-1984}.
This index is often used to model stellar activity in RV time-series \citep[e.g.][]{Dumusque-2011c,Diaz:2016ab} or to train non-parametric models like Gaussian Processes to mitigate stellar activity \citep[e.g.][]{Rajpaul-2015, Langellier:2020aa}.

\begin{figure*}
        \center
	\includegraphics[width=18cm]{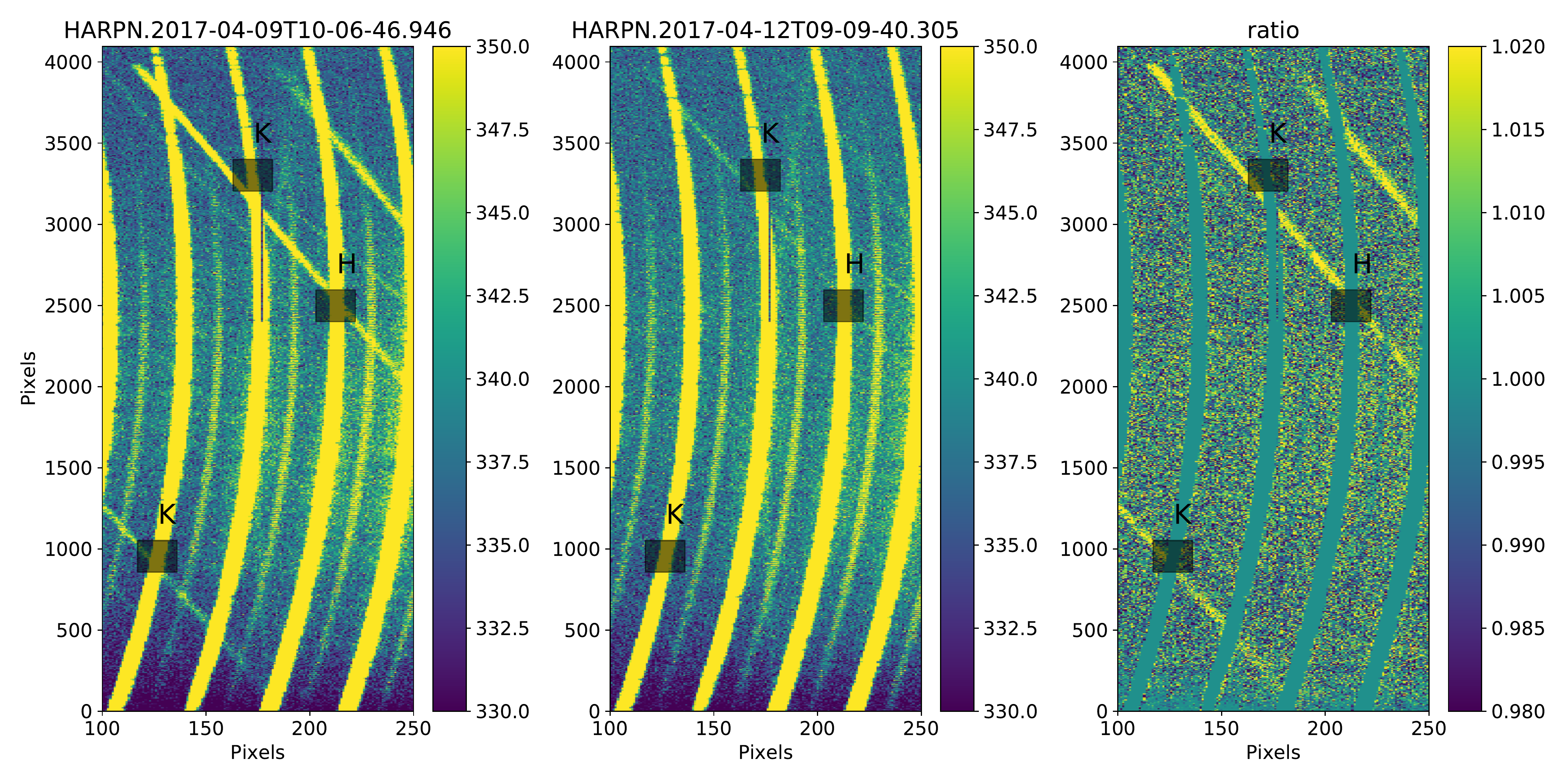}
	\caption{Raw images from HARPS-N taken before (HARPN.2017-04-09T10-06-46.946, left panel) and after (HARPN.2017-04-12T09-09-40.305, middle panel) a warm-up of the detector, and their ratio (right panel). The location of the core of the Ca II K and H lines in orders one, two and three is highlighted by grey boxes. As we can see, the core of the Ca II K line in order one (K, bottom left) and H line in order three (H) are strongly contaminated by ghosts. In addition, we also see that the contamination due to ghosts is significantly reduced with the warm-up process, which removes the humidity accumulated over time on the detector due to a small leak in its cryostat.}
	\label{fig:ghost_contam}
\end{figure*}

However, the S-index or \logrhk\,is difficult to measure as the related chromospheric emission appears in the extreme blue. At those wavelengths, more absorption from the Earth's atmosphere, and lower sensitivity of optical fibres, optics and detector make it difficult to reach high S/N. In addition, the chromospheric emission happens in the core of the extremely saturated Ca II H and K spectral lines, where the S/N is even lower. Therefore, a careful background correction has to be performed to extract correctly the chromospheric emission. In addition, on HARPS-N, a small leak in the detector cryostat causes humidity to build-up on the detector, which increases its reflectivity over time. A periodic warm-up of the detector solves the issue for a little while. However, with the reflectivity increase, the brightness of ghosts increases as well, modifying the contamination in the extreme blue with time. On the first and second panels of Fig.~\ref{fig:ghost_contam}, we show the bluest part of two raw solar science frames of similar S/N obtained with HARPS-N (frames HARPN.2017-04-09T10-06-46.964.fits and HARPN.2017-04-12T09-09-40.30.fits), with the Sun on the science fibre and the Fabry-P\'erot on the reference fibre. The images have been taken just before and after a warm-up of the detector, and show how the ghosts significantly reduce in intensity with the intervention.

\begin{figure*}
        \center
	\includegraphics[width=12.93cm]{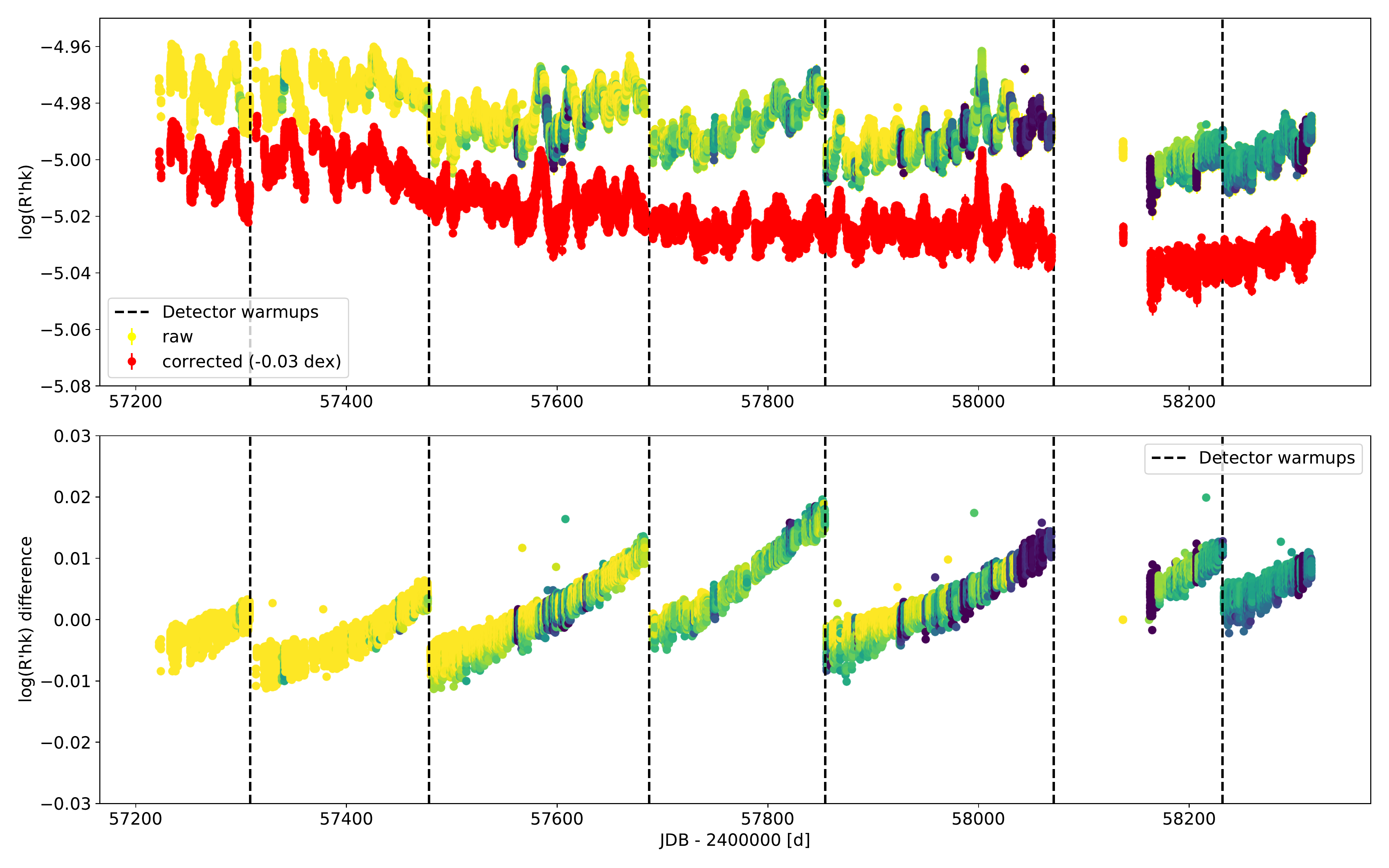}
	\includegraphics[width=5.38cm]{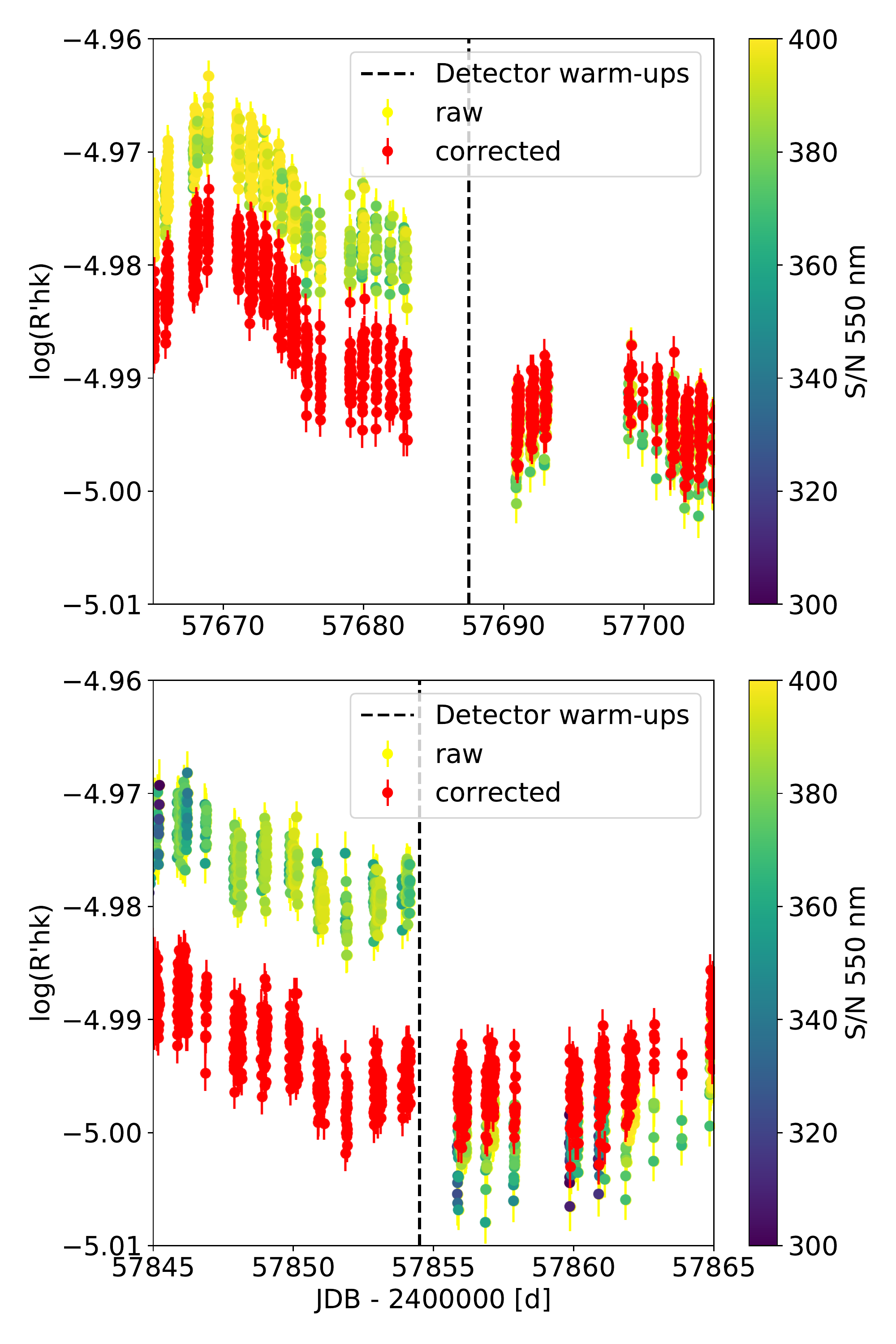}
	\caption{\emph{Top left: }Comparison between the raw (dots with yellow to blue color scale) and corrected (red dots) \logrhk\,calcium activity index, as described in the text. \emph{Bottom left: } Differences between the raw and corrected \logrhk\,time-series. We clearly see that the slow increase in activity index followed by sudden jumps at each warm-up of the detector seen in the raw \logrhk\,time-series is strongly reduced after correction. \emph{Top and bottom right:} Closer look at the regions around the third and fourth warm-ups.}
	\label{fig:rhk}
\end{figure*}

In Fig.~\ref{fig:ghost_contam}, we highlighted the location of the Ca II H and K lines. We note that the core of the Ca II K line, appears on two consecutive orders, extracted orders one and two, and the core of the Ca II H lines appears on the third extracted order. The core of the K line in order one and the H line in order three are significantly affected by ghosts. With humidity building up on the detector over time, ghost contamination increases, and therefore increases the measured \logrhk. After a warm-up, the contamination is significantly reduced, which induces an abrupt offset in the \logrhk\,time-series. This is clearly visible in the top left panel of Fig.~\ref{fig:rhk}, where we show the raw \logrhk\,extracted by the old HARPS-N DRS along with the time-localisation of each warm-up.

The old and new HARPS-N DRS do not correct for ghost contamination, as this is challenging to perform. First of all, ghosts are at low S/N which makes their extraction difficult. In addition, ghosts are oblique relative to echelle orders and contaminate them over hundreds of pixels. Finally ghosts, which are due to secondary reflection of the echelle orders, do not have a smooth varying flux, but the flux of the observed solar spectrum. Because it is difficult to know which ghost corresponds to which echelle order, we decided to use a rather simple technique, described in Appendix~\ref{app:rhk_correction}, to correct for the contamination.

The \logrhk\,measured with and without the correction can be seen in the top left panel of Fig.~\ref{fig:rhk} and the difference between the two is shown in the bottom left panel. It is clear that the increase of flux over time followed by sudden jumps at each detector warm-up is strongly mitigated in the corrected time-series. A closer look at the regions around the third and fourth warm-ups can be seen in the right panel of the same figure. For the third warm-up, the observed offset without correction is 0.015 dex, and it goes down to 0.003 dex after correction. For the fourth warm-up the offset goes from 0.024 to 0.002 dex. The correction applied therefore mitigate strongly the effect of ghost contamination in the \logrhk time-series, and it is difficult to say at this stage if the tiny offset remaining at the third and fourth warm-ups are from solar or from instrumental origin.

The corrected \logrhk\,activity index that we can see in the top left panel of Fig.~\ref{fig:rhk} has an exquisite precision due to the high S/N solar observations. The median error bar is 0.0014 dex for 5-minute integration times, with a standard deviation of 0.0002 dex.
As we can see, the 25-day rotational modulation due to active regions turning with the Sun is clearly visible. Detailed analysis of this \logrhk\,time-series was performed in \citet[][]{Maldonado:2019aa} and \citet[][]{Milbourne:2019aa}. We note that in those papers, the \logrhk\,time-series were corrected with a similar method as described here.

\section{The HARPS-N solar data set} \label{RV_data_set}

\subsection{Data selection} \label{data _selection}
For this paper, and the related data release, we
only selected good quality spectra over the first three years of operation of the HARPS-N solar telescope. As the Sun is resolved with the 3 inch entrance lens of the solar telescope, any kind of absorption,
such as clouds, not evenly distributed over the solar disc will induce a huge Rossiter-McLaughlin-type effect, with RVs that can depart from the true value by several 
hundreds of meters-per-second, as the $\vsini$ of the Sun is 2\kms. To determine if spectra are contaminated by clouds, \citet{Collier-Cameron2019aa} performed a daily regression of $-2.5\log[(S/N)^2]$ in echelle order 60 (as a proxy for apparent solar magnitude) against airmass. Outliers from the expected linear extinction relation each day were considered to have been affected by clouds, and were assigned probabilities of belonging to a separate outlier population using a Gaussian mixture model \citep[][]{Foreman-Mackey-2014aa, Hogg:2010aa}. For the present paper, and the related data release, we only considered the data for which the probability of being of good quality is $\ge0.99$.

With such a threshold, 65\% (34550) of the spectra taken with the HARPS-N solar feed are selected out of the 53417 available. The median cadence between consecutive observations, excluding the night gaps, is 5.42 minutes, which is consistent with the effective time between measurements; 5 minutes exposure time plus 20 seconds readout. We note that we chose an exposure of 5 minutes to average out the signals induced by p-mode oscillations \citep[e.g.][]{Chaplin:2019aa}.

\subsection{Comparison between the solar radial velocities reduced with the new and old HARPS-N DRS} \label{RV_comparison}

In this section, we compare the three years of solar RVs already published in \citet{Collier-Cameron2019aa}, obtained with the old HARPS-N DRS and referred to as the old solar RVs, with the RVs obtained with the ESPRESSO DRS and the improvements discussed in this paper (i.e. the new HARPS-N DRS), referred to as the new solar RVs. When estimating RV errors, the old HARPS-N DRS was including photon-noise, drift noise induced by the drift measurement performed on the simultaneous Fabry-P\'erot spectrum and the error in wavelength calibration, extracted from the polynomial fit performed to derive it. The median value for this error in wavelength calibration is 0.35\ms, which is much less than the 0.74\ms measured in Sect.~\ref{sect:new_wave_sol}. Thus, the HARPS-N RVs derived using the old DRS have error bars that are significantly underestimated.

In the new DRS, the estimated RV errors only include photon and drift noises, because to account properly for the wavelength solution noise, which impact in the same way all the observations performed within the same calibration set (obtained on average every day), we need to consider a covariance matrix which is block-diagonal; each block regrouping all the observations attached to the same wavelength solution \citep[][]{Delisle:2020ab}. Each block should thus have the same calibration noise of 0.49\ms for the new DRS data. Another way to include correctly this calibration noise is to daily-bin the RV data, and then add in quadrature a calibration noise of 0.49\ms to the RV errors.

\begin{figure*}
        \center
	\includegraphics[width=19cm]{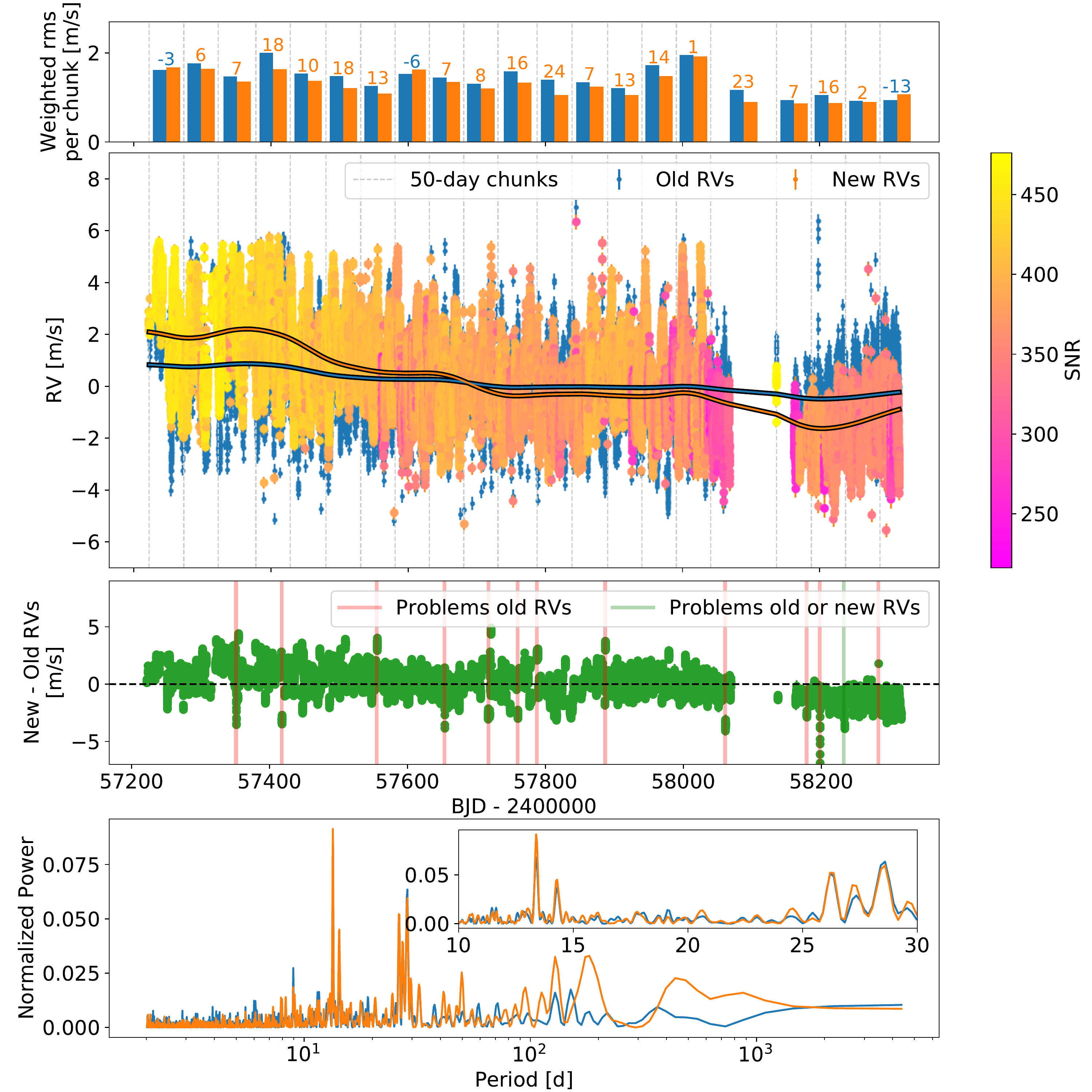}
	\caption{Comparison between the solar RVs reduced using the old HARPS-N (blue) and the new HARPS-N DRS (orange). \emph{Top panel:} Weighted rms for each chunk of RV data of 50 days, which corresponds to two solar rotations and therefore include activity signal. Orange bars correspond to the weighted rms measured on the new RVs, and blue bars on the old RVs. The number for each chunk corresponds to the improvement in weighted rms brought by new DRS (in percentage). \emph{Second panel:} Solar RVs as derived by the old and new HARPS-N DRS. The blue and orange lines correspond to a smooth version of the calcium activity index variation linearly fitted to the RVs, to remove the strong long-term correlation between RVs and the \logrhk, and thus to mitigate the contribution from the solar magnetic cycle (see text). The color scale for the new RVs corresponds to the S/N of the measurements. We note that we lost a significant amount of data around BJD=2458100 (December 2017), due to a damaged fibre (see text). \emph{Third panel:} Difference between the new and old solar RVs, for which a long-term trend is clearly visible. We highlighted with red and green vertical lines the significant discrepancies in those RV differences. The red lines corresponds to problems in the old DRS RVs, while for the green vertical line, it is not clear if the correct RVs are the ones derived with the old or new DRS. \emph{Bottom panel:} Generalized Lomb-Scargle Periodogram of the residuals after removing the long-term trend observed in the RVs. The small inset shows a zoom around the periods 10 to 30 days to highlight stellar rotation.}
	\label{fig:rv_sun}
\end{figure*}

In the second panel of Fig.~\ref{fig:rv_sun}, we show the comparison between the old and new solar RVs. The color scale for the new RVs corresponds to the S/N of the measurements. As we can see, the S/N degraded from $\sim$460 to $\sim$350 in three years, thus a reduction in flux of nearly a factor of two. We never changed the exposure time nor the neutral density filter used to attenuate the incoming flux, thus this could be due to ageing of the solar telescope plexiglass dome, the optics of the solar telescope or the fibre guiding solar light into HARPS-N or a mix of the three. The neutral density filter was changed recently to solve this issue. We note that around BJD=2458100 (December 2017), the fibre guiding the light from the solar telescope to HARPS-N was damaged following an intervention on the HARPS-N calibration unit. This led to observations with very low S/N, giving RVs that were not precise enough for this study. Those data were rejected from the data released, explaining the absence of data around BJD=2458100.

\begin{figure*}
        \center
	\includegraphics[width=18cm]{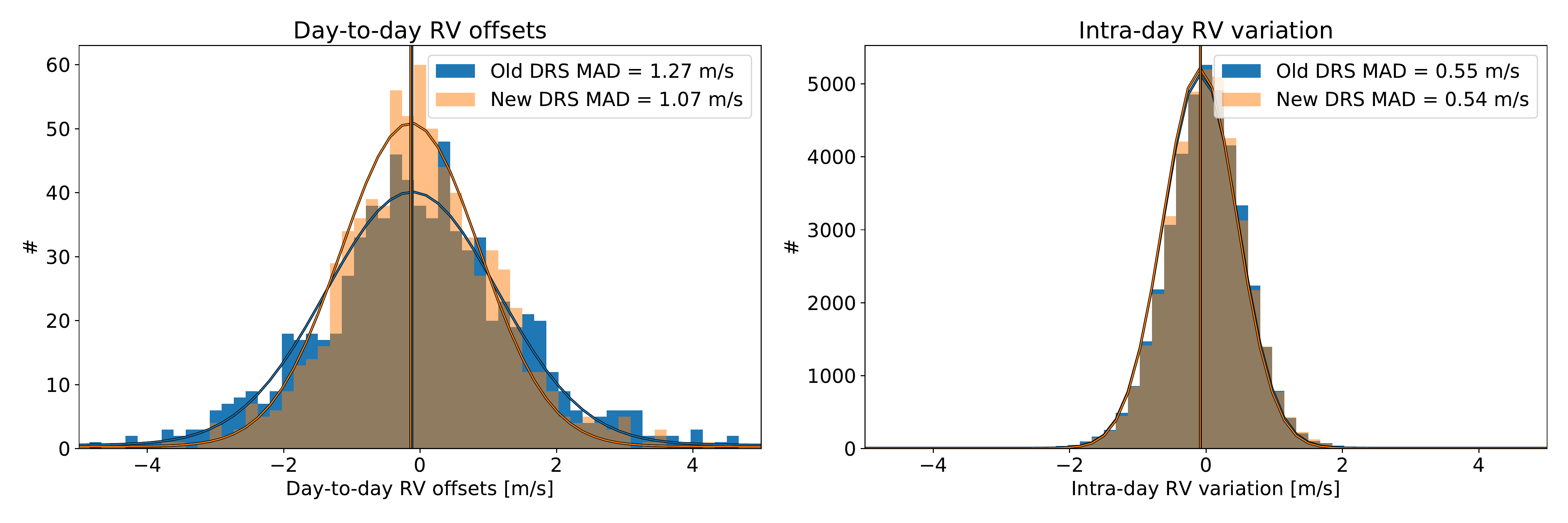}
	\caption{ \emph{Left:} Histogram of the day-to-day RV offsets measured on the old and new solar RVs after daily-binning the data. \emph{Right:} Intra-day RV variation obtained by removing the daily mean on each single day.}
	\label{fig:dayBday_offsets}
\end{figure*}

To see the improvements brought by the new DRS, we divided the data set in chunks of 50 days, which corresponds to approximately two solar rotations, and measured the weighted rms within each chunk (values reported in the top panel of Fig.~\ref{fig:rv_sun}). We find that 17 of the 21 chunks show a greatly reduced weighted rms in the new solar RVs, with an increased precision of up to 24\%. This improvement is mainly brought by i) the new drift map method with tailored thorium line selection to derive wavelength solutions (see Sect.~\ref{sect:new_wave_sol}), which reduces calibration to calibration RV offsets by 0.55\ms, and ii) by the use of the afternoon calibrations rather than the ones from the previous day. In the left panel of Fig.~\ref{fig:dayBday_offsets}, we show the day-to-day RV offsets measured on the old and new RVs after daily-binning the data. As we can see, the new RVs present a distribution with a much smaller MAD of 1.07\ms, compared to 1.27\ms for the old RVs. Thus, when analysing day-to-day RV offsets, the new solar RVs show a very significant improvement of $\sqrt{1.27^2-1.07^2}=0.68$\ms. 

In the right panel of Fig.~\ref{fig:dayBday_offsets}, we show the RV variation on timescales shorter than a day, which is obtained by subtracting the daily mean of each day. On this short timescale, we measure a very similar MAD for the new and old DRS, 0.54 and 0.55\ms, respectively, which is not surprising as all the data within a day generally use the same set of calibrations. This MAD is significantly larger than the median photon noise of 0.25\ms. This discrepancy is likely due to stellar granulation \citep[e.g.][]{Meunier:2017ab, Dumusque-2011a}, and oscillations that are not fully averaged-out with an integration time of 5 minutes \citep[e.g.][]{Chaplin:2019aa}, as we know that those stellar signals are of several dozens of meters-per-second on the Sun \citep[e.g.][]{Lefebvre-2008}.

In the third panel of Fig.~\ref{fig:rv_sun}, we can see the difference in RVs between the new and old solar RVs. We see that the two data sets follow each other closely, except for a few points highlighted with red and green vertical lines. After visually inspecting in the old and new DRS RVs if the corresponding data points were following the general RV variation or if they were outliers, we concluded that the 12 red vertical lines corresponds to discrepancies induced by the old DRS. The one green line corresponds to data for which it is difficult to say if the old or the new DRS gives a better solution.

\begin{figure}
        \center
	\includegraphics[width=9cm]{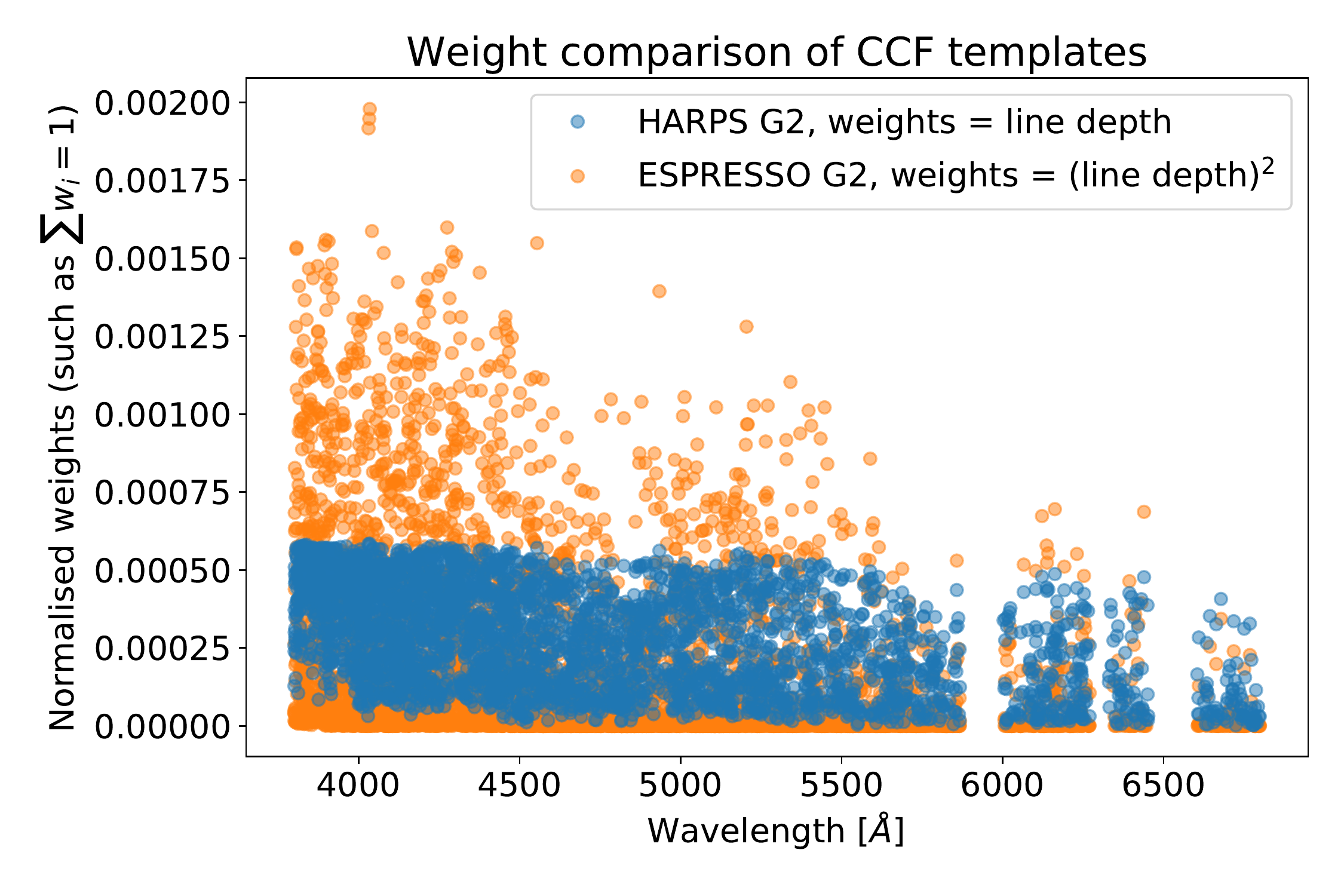}
	\caption{ Comparison of the weights for all the lines used to build the CCF between the old and new DRS G2 templates.}
	\label{fig:compa_w_CCF_masks}
\end{figure}

Another significant aspect that can be seen in the RV differences is a small long-term drift rather linear with time. Except probably for the behavior observed after BJD=2458070, this drift does not correlate well with the difference in wavelength solutions between the old and new DRS (see right panel of Fig.~\ref{fig:th_lines_used_and_wavesol_compa}). Therefore, it does not seem that it comes from the way wavelength solutions are derived. One thing that is different between the old and new DRS, is the computation of the cross-correlation function \citep[CCF,][]{Baranne-1996}. First, the old reduction uses the HARPS-N G2 template to perform the cross-correlation, while the new DRS uses the ESPRESSO G2 template. Those templates consist of the wavelength of all the spectral lines used to compute the CCF, with some weights corresponding to the strength of the lines. As described in \citet{Pepe-2002a}, lines that are stronger have a better RV precision and therefore should be weighted more when computing the CCF. In the old DRS, the CCF is computed using as weight the normalized depth of the lines with respect to the continuum, as in \citep[][]{Pepe-2002a}, while the new DRS uses the square of the line depth, which corresponds better to the true RV precision of a spectral line \citep[][]{Lovis:2010aa}. As lines in the blue part of the spectrum of a G2 star are very strong and their density is higher than in the red, using as weight the depth of the lines squared put more importance on the blue part of the spectrum, as shown in Fig.~\ref{fig:compa_w_CCF_masks}. Several studies demonstrate on simulations that the blue part of the visible is more sensitive to stellar activity \citep[e.g.][]{Reiners-2010, Desort-2007}. This has also been shown on real data, but generally on stars more active than the Sun \citep[][]{Zechmeister:2018aa, Feng:2017ac, Anglada-Escude-2012}. If it is the case for the Sun, we expect the new solar RVs to be more sensitive to stellar activity than the old ones.

\begin{figure}
        \center
	\includegraphics[width=9cm]{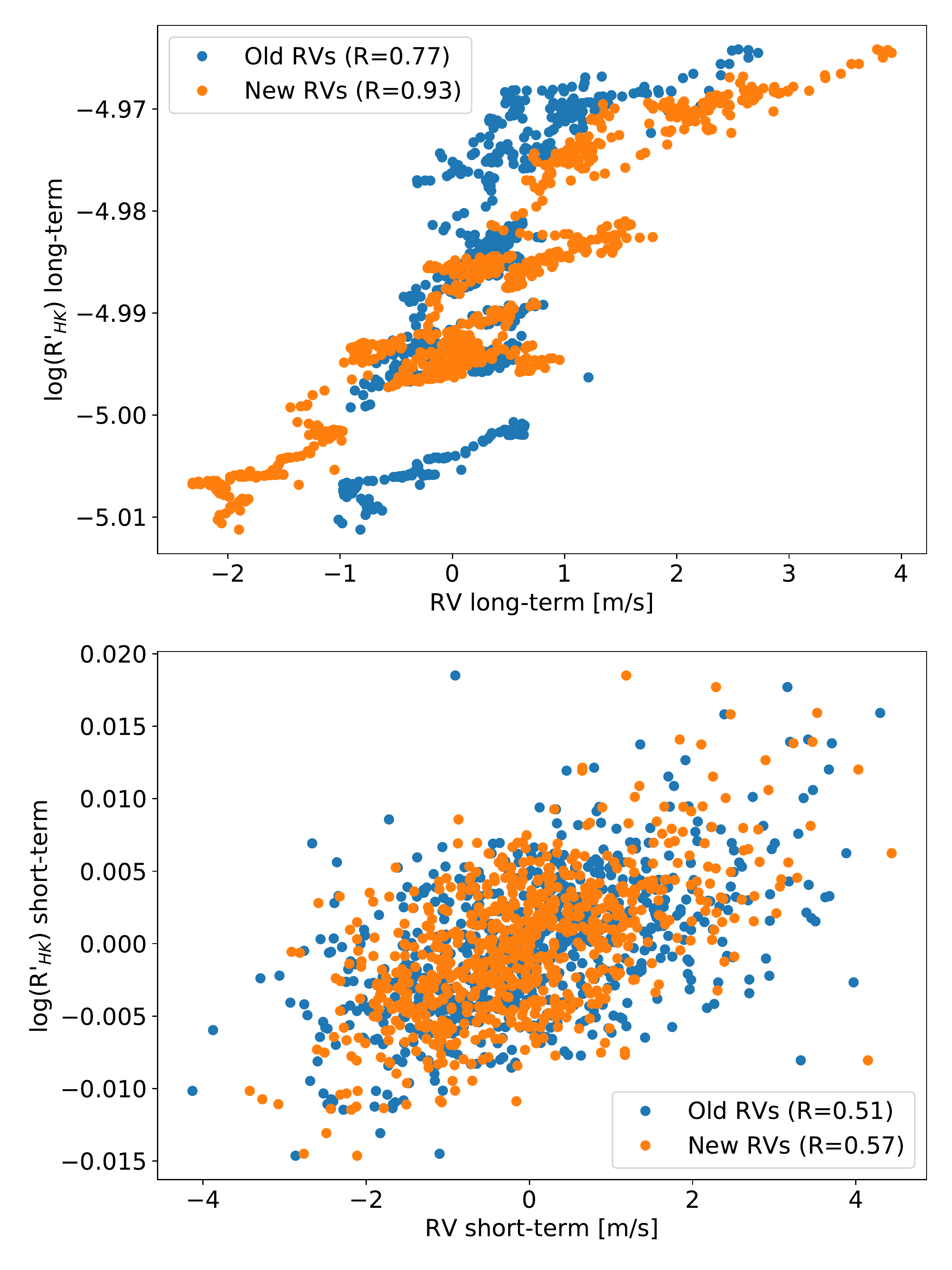}
	\caption{\emph{Top: }Correlation between the RVs and the \logrhk\,for long-term variations, which in this case means that the time-series have been smoothed using a 30-day rolling average. \emph{Bottom: }Correlation between the short-term variations, meaning the raw time-series minus the smoothed ones. We show in the legend the values for the R Pearson correlation coefficient for the new and old solar RVs.}
	\label{fig:rv_rhk_corr}
\end{figure}

On stars like the Sun, the main source of activity is induced by faculae \citep[e.g.][]{Meunier-2010b,Haywood-2016,Milbourne:2019aa}. The magnetic field in faculae inhibits locally stellar convection, which suppress locally the convective blueshift and therefore induces a net redshift \citep[e.g.][]{Lindegren-2003, Meunier-2010b}. Thus, it is expected that the calcium activity index, probing faculae, and the RV exhibit a strong correlation \citep[e.g.][]{Lindegren-2003, Lovis-2011b, Dumusque-2011c, Meunier:2017aa}. By looking at the correlation between the RV and the \logrhk\,time-series in Fig.~\ref{fig:rv_rhk_corr}, we clearly see that on the long-term, the new solar RVs show a stronger correlation, with a Pearson correlation coefficient of $R=0.93$ compared to $R=0.77$ for the old RVs. Regarding short-term variations, the new solar RVs correlate slightly better, with $R=0.57$ compared to $R=0.51$, however, the difference is less significant. 
This analysis confirms that the new solar RVs are more sensitive to stellar activity than the old ones, and as explained in the preceding paragraph, this is likely because the new DRS uses a CCF template that puts more weight on the blue part of the spectrum.

In the bottom panel of Fig.~\ref{fig:rv_sun}, we compared the Generalized Lomb-Scargle periodograms \citep[][]{Zechmeister-2009} of the new and old solar RVs. As it seems that the long-term drift is induced by the solar magnetic cycle, we removed its contribution by linearly fitting to the RVs a smooth version of the calcium activity index variation\footnote{We note that the smooth version of the calcium activity index was obtained by applying the low-pass filter algorithm described in \citet{Press-1989} and by selecting a 200 days$^{-1}$ cutoff frequency}, as performed in several studies \citep[e.g.][]{Delisle:2018aa, Diaz:2016ab}. The best-fitted linear models are represented as smooth lines in the second panel of the same figure. At a first glance, the two periodograms are nearly identical. As we can see, the most significant signals can be found around 27 and 13.5 days, the synodic rotational period of the Sun, and its first harmonic, which is expected from stellar activity \citep[e.g.][]{Saar-1997b, Boisse-2009}. In the small inset in the bottom panel of the figure, we clearly see that those signals modulated by stellar rotation have more power in the new DRS RVs, mainly for the peak at 13.4 days. This can be explained by the fact that the new solar RVs are less affected by instrumental and data reduction systematics, and therefore the signal from stellar activity is better characterized. We note however, a slight excess of power around 180 days in the new DRS RVs. This periodicity is also observed in independent observations of the Sun over magnetic cycle 24, estimated using images from the Helioseismic and Magnetic Imager onboard the Solar Dynamics Observatory, in the RV variations as well as in the plage filling factor and unsigned magnetic flux \citep[see Fig.~5 in][]{Haywood:2020aa}. A periodicity of $\sim$155 days is also found in the sunspot blocking function, the 10.7 cm radio flux and the sunspot number of solar cycle 19, 20 and 21 \citep[][]{Lean:1989aa} and in the recurrence of hard solar flares in cycle 21 \citep[][]{Rieger:1984aa}. We conclude that this 180-day periodicity is probably from stellar origin and not from instrumental systematic, however further investigation is required.

\subsection{Data release} \label{data_release}

Alongside with this paper, we release the first three years of HARPS-N solar data. Those data consist of the following data products reduced with the new HARPS-N pipeline:
\begin{itemize}
\item The extracted echelle-order spectra, corrected from the instrumental blaze, in the Earth rest-frame. These products are called S2D spectra due to their two dimensional shape.

\item The extracted merged spectra, corrected from the instrumental blaze, in the Solar System barycentric rest-frame. These products are called S1D spectra.

\item The CCFs obtained by cross-correlating the S2D spectra with a synthetic mask derived from solar observations (ESPRESSO G2 mask), thus optimized for the Sun.
\end{itemize}
In addition to those data products, we also release the time-series of relevant quantities such as the RV, the BIS SPAN, the FWHM, different activity indicators. They are all listed in Table~\ref{tab:temeseries_list} with their descriptions.

The different data products as well as the time-series can be accessed using the Data Analysis Center for Exoplanet (DACE) web-platform at this address \url{https://dace.unige.ch/sun/}, or using the DACE python API (\url{https://dace.unige.ch/pythonAPI/} with the solar spectroscopy module). More detailed about the available data and how to access them can be found in Appendix \ref{app:data_release}.

\section{Discussion and conclusion} \label{Conclusion}

In this paper, we describe a new data reduction for HARPS-N that allowed us to obtain unprecedented RV precision for the three years of solar data that are released alongside this manuscript (see Sect.~\ref{data_release} and Appendix~\ref{app:data_release} for more information about the released products). We also compared the newly derived RVs with the ones already published in \citet{Collier-Cameron2019aa}, and clearly demonstrated that the new data are less affected by instrumental and data reduction systematics.

Compared to the \citet{Collier-Cameron2019aa} solar RVs data that were reduced using the current HARPS-N DRS (version 3.7, referred to as the old DRS), we reduced new RVs using the ESPRESSO DRS tailored to HARPS-N (version 2.2.3, referred to as the new DRS, see Sect.~\ref{espresso_drs}). In addition, we developed a new method to derive wavelength solutions (see Sect.~\ref{sect:new_wave_sol} and Appendix~\ref{app:drift_map_method}) and curated the list of thorium lines that are used when deriving a wavelength solution, so that only the most stable thorium lines over time are used (see Sect.~\ref{sect:th_line_selection} and Appendix~\ref{app:th_line_selection}). We also reduced the data taking the closest calibrations in the future rather than in the past (see Sect.~\ref{sect:afternoon_calib}) to mitigate any systematics induced by Fabry-P\'erot internal drifts, and we used master flat fields that have a much higher S/N than standard flat fields, to prevent the solar observations being limited by them (see Sect.~\ref{sect:master_flat_field}). We also extracted a calcium activity index, which has been corrected from ghost contamination (see Sect.~\ref{correcting_calcium} and Appendix~\ref{app:rhk_correction}).

We performed several analysis in Sect.~\ref{RV_comparison} to compare the new and old DRS solar RVs. When dividing the solar data in chunks of 50 days to still see the 25-day effect of stellar rotation, we observe that for most of the chunks, the new solar RVs are less affected by noise, with improvement of up to 24\% (see Fig.~\ref{fig:rv_sun}). At first, we could think that the new DRS mitigate part of the stellar activity signal. We will see below, when discussing the correlation between the RVs and the activity index \logrhk\, that this does not seem to be the case. In fact, a similar improvement is seen when studying the RV variability on a timescale of a day, which is more affected by instrumental and data reduction systematics as calibrations are performed on a daily basis. On that timescale, it is clear from the left panel of Fig.~\ref{fig:dayBday_offsets} that the new RVs are less affected by noise as the measured MAD goes down to 1.07\ms with the new DRS, compared to 1.27\ms with the old one. As systematics adds up in quadrature, this implies that the new reduction reduces a noise of 0.68\ms present in the old solar RVs. This improvement is mainly brought by the new wavelengths solution derivation method, the optimal selection of thorium lines, and the use of calibrations closer in time. 

A noise floor of 1.07\ms on a daily basis is still much larger than the photon noise error, of a few centimeters-per-second\footnote{For a single day, with a median photon noise error of 0.25\ms and a median number of observation per day of 56, the estimated photon-noise error is $\sim0.25/\sqrt{56}$=0.03\ms}., and the wavelength calibration noise error, estimated to be 0.49\ms when analysing the RV offsets between consecutive wavelength solutions (See Sect.~\ref{sect:new_wave_sol} and Fig.~\ref{fig:inst_drift}). Although we cannot exclude some remaining instrumental and data reduction systematics, it is likely that the remaining 0.96\ms ($\sqrt{1.07^2-0.03^2-0.49^2)}$) in RV variation over a day are induced by stellar signals. On that timescale, several studies have shown that super-granulation on solar-type stars can induce a RV variability at the level of the meter-per-second \citep[e.g.][]{Lefebvre-2008, Dumusque-2011a, Meunier-2015}.

We also looked at the RV variability within a day. In the right panel of Fig.~\ref{fig:dayBday_offsets}, we see that the old and new data reductions gives a similar MAD of 0.55 and 0.54\ms, respectively. As all data within a day are generally reduced with a single calibration set, and assuming that the Fabry-P\'erot \'etalon that is used to correct for instrumental drift does not drift internally, we do expect to measure a RV variability close to photon-noise. As for a single exposure, the median photon-noise is 0.25\ms, the solar data are clearly affect by some extra variability. This noise is likely due to stellar granulation \citep[e.g.][]{Meunier:2017ab, Dumusque-2011a}, and oscillations that are not fully averaged-out with an integration time of 5 minutes \citep[e.g.][]{Chaplin:2019aa}, as we know that those stellar signals are of several dozens of meters-per-second on the Sun \citep[e.g.][]{Lefebvre-2008}.

We observed that the new solar RVs show a stronger drift with time than the old RVs (see Fig.~\ref{fig:rv_sun}), and this drift cannot be explained by the use of the new selection of thorium lines and drift map wavelength solutions (see Sect.~\ref{RV_comparison}). The presence of this stronger drift induces a stronger correlation of $R=0.93$, compared to $R=0.77$, between the RVs and the \logrhk\,on the long-term (see Fig.~\ref{fig:rv_rhk_corr}), which is expected in the case of a magnetic cycle as seen on the Sun \citep[e.g.][]{Dumusque-2011c, Meunier-2010a,Lindegren-2003}. After investigation, it seems that this drift is mainly induced by the use in the new DRS of a new cross-correlation template, the ESPRESSO G2 mask, that put more weight on the blue part of the spectrum. The blue is more sensitive to stellar activity effects \citep[e.g.][]{Zechmeister:2018aa, Feng:2017ac, Anglada-Escude-2012, Reiners-2010, Desort-2007}, and thus the RVs are more affected by the solar magnetic cycle. We believe that being more sensitive to stellar activity can be seen as an improvement, as this increase the S/N of the activity signal, which helps techniques that needs to better characterize stellar activity before mitigating it \citep[e.g.][]{Rajpaul-2015, Cretignier:2020aa, Collier-Cameron:2020aa, Beurs:2020aa}. We however have to keep in mind that this will complicate the search for planetary signals if no stellar activity correction is performed.

On the timescale of solar rotation, the old and new DRS gives a similar correlation of $R=0.51$ and $R=0.57$. This confirms that although the \logrhk\,can be used to efficiently model the RV variations induced by solar-like magnetic cycles \citep[e.g.][]{Lovis-2011b, Meunier:2020aa}, it cannot be used to model efficiently the RV stellar activity signals on the rotational timescale.

From the different analysis performed in this paper, it is clear that the new solar RVs and corresponding high-resolution spectra and CCFs should be preferred compared to the old DRS products, as the instrumental and data reduction RV variability is smaller, but also because those data better reflect the effect from the solar magnetic cycle. After releasing these data through the DACE web-platform (see end of Sect.~\ref{RV_data_set}), the next effort from the HARPS-N collaboration will be to run the new DRS with all the described improvements in online mode at the Telescopio Nazionale Galileo, and to re-reduce all the HARPS-N data, as the gain in term of precision and stability is clearly demonstrated on the Sun.

On a final note, we reduced at best all the observed systematics on the HARPS-N solar data set. The release of this unprecedented time-series of spectra and radial velocities is crucial to improve our understanding of stellar signals affecting the Sun and solar-type stars. We hope that the community will use such data to develop novel methods to mitigate stellar signals in radial-velocity data sets, with the goal of enabling the detection of other Earths.

\begin{acknowledgements}
XD is grateful to The Branco Weiss Fellowship--Society in Science for its financial support. 
This project has received funding from the European Research Council (ERC) under the European Union's Horizon 2020 research and innovation programme (grant agreement No 851555/SCORE).
ACC acknowledges support from
the Science and Technology Facilities Council (STFC) consolidated
grant number ST/R000824/1.
This work has been carried out in the framework of the National Centre for Competence in Research \emph{PlanetS} supported by the Swiss National Science Foundation (SNSF). AM acknowledges support from the senior Kavli Institute Fellowships. 
This research used the DACE platform developed in the frame of PlanetS (\url{https://dace.unige.ch}).
This work was performed under contract with the California Institute of Technology (Caltech)/Jet Propulsion Laboratory (JPL) funded by NASA through the Sagan Fellowship Program executed by the NASA Exoplanet Science Institute (RDH).
\end{acknowledgements}

\bibliographystyle{aa}
\bibliography{dumusque_bibliography}

\begin{thebibliography}{63}
\expandafter\ifx\csname natexlab\endcsname\relax\def\natexlab#1{#1}\fi

\bibitem[{{Anglada-Escud{\'e}} \& {Butler}(2012)}]{Anglada-Escude-2012}
{Anglada-Escud{\'e}}, G. \& {Butler}, R.~P. 2012, \apjs, 200, 15

\bibitem[{{Baliunas} {et~al.}(1995){Baliunas}, {Donahue}, {Soon}, {Horne},
  {Frazer}, {Woodard-Eklund}, {Bradford}, {Rao}, {Wilson}, {Zhang}, {Bennett},
  {Briggs}, {Carroll}, {Duncan}, {Figueroa}, {Lanning}, {Misch}, {Mueller},
  {Noyes}, {Poppe}, {Porter}, {Robinson}, {Russell}, {Shelton}, {Soyumer},
  {Vaughan}, \& {Whitney}}]{Baliunas-1995}
{Baliunas}, S.~L., {Donahue}, R.~A., {Soon}, W.~H., {et~al.} 1995, \apj, 438,
  269

\bibitem[{{Baranne} {et~al.}(1996){Baranne}, {Queloz}, {Mayor}, {Adrianzyk},
  {Knispel}, {Kohler}, {Lacroix}, {Meunier}, {Rimbaud}, \&
  {Vin}}]{Baranne-1996}
{Baranne}, A., {Queloz}, D., {Mayor}, M., {et~al.} 1996, \aaps, 119, 373

\bibitem[{{Bauer} {et~al.}(2015){Bauer}, {Zechmeister}, \&
  {Reiners}}]{Bauer:2015aa}
{Bauer}, F.~F., {Zechmeister}, M., \& {Reiners}, A. 2015, \aap, 581, A117

\bibitem[{{Blackwood} {et~al.}(2020){Blackwood}, {Gaudi}, {Burt}, {Mamajek},
  {Beichman}, {Cegla}, {Fischer}, {Ford}, {Howard}, {Latham}, {Plavchan},
  {Quirrenbach}, \& {EPRV Working Group}}]{eprvwg}
{Blackwood}, G., {Gaudi}, B.~S., {Burt}, J., {et~al.} 2020, in American
  Astronomical Society Meeting Abstracts, American Astronomical Society Meeting
  Abstracts, 374.01

\bibitem[{{Boisse} {et~al.}(2009){Boisse}, {Moutou}, {Vidal-Madjar}, {Bouchy},
  {Pont}, {H{\'e}brard}, {Bonfils}, {Croll}, {Delfosse}, {Desort}, {Forveille},
  {Lagrange}, {Loeillet}, {Lovis}, {Matthews}, {Mayor}, {Pepe}, {Perrier},
  {Queloz}, {Rowe}, {Santos}, {S{\'e}gransan}, \& {Udry}}]{Boisse-2009}
{Boisse}, I., {Moutou}, C., {Vidal-Madjar}, A., {et~al.} 2009, \aap, 495, 959

\bibitem[{{Cersullo} {et~al.}(2018){Cersullo}, {Coffinet}, {Chazelas}, {Lovis},
  \& {Pepe}}]{Cersullo-2018}
{Cersullo}, M.~F., {Coffinet}, A., {Chazelas}, B., {Lovis}, C., \& {Pepe}, F.
  2018, \aap, submitted

\bibitem[{{Chaplin} {et~al.}(2019){Chaplin}, {Cegla}, {Watson}, {Davies}, \&
  {Ball}}]{Chaplin:2019aa}
{Chaplin}, W.~J., {Cegla}, H.~M., {Watson}, C.~A., {Davies}, G.~R., \& {Ball},
  W.~H. 2019, \aj, 157, 163

\bibitem[{{Claudi} {et~al.}(2018){Claudi}, {Ghedina}, {Pace}, {Gallorini}, {Di
  Giorgio}, {Liu}, {Tozzi}, {Lanza}, {Micela}, {Molinari}, {Phillips}, \&
  {Tripodo}}]{Claudi:2018aa}
{Claudi}, R., {Ghedina}, A., {Pace}, E., {et~al.} 2018, in Society of
  Photo-Optical Instrumentation Engineers (SPIE) Conference Series, Vol. 10700,
  Ground-based and Airborne Telescopes VII, 107004N

\bibitem[{{Collier Cameron} {et~al.}(2020){Collier Cameron}, {Ford}, {Shahaf},
  {Aigrain}, {Dumusque}, {Haywood}, {Mortier}, {Phillips}, {Buchhave},
  {Cecconi}, {Cegla}, {Cosentino}, {Cretignier}, {Ghedina}, {Gonzalez},
  {Latham}, {Lodi}, {Lopez-Morales}, {Micela}, {Molinari}, {Pepe}, {Piotto},
  {Poretti}, {Queloz}, {San Juan}, {Segransan}, {Sozzetti}, {Szentgyorgyi},
  {Thompson}, {Udry}, \& {Watson}}]{Collier-Cameron:2020aa}
{Collier Cameron}, A., {Ford}, E.~B., {Shahaf}, S., {et~al.} 2020, arXiv
  e-prints, arXiv:2011.00018

\bibitem[{{Collier Cameron} {et~al.}(2019){Collier Cameron}, {Mortier},
  {Phillips}, {Dumusque}, {Haywood}, {Langellier}, {Watson}, {Cegla}, {Costes},
  {Charbonneau}, {Coffinet}, {Latham}, {Lopez-Morales}, {Malavolta},
  {Maldonado}, {Micela}, {Milbourne}, {Molinari}, {Saar}, {Thompson},
  {Buchschacher}, {Cecconi}, {Cosentino}, {Ghedina}, {Glenday}, {Gonzalez},
  {Li}, {Lodi}, {Lovis}, {Pepe}, {Poretti}, {Rice}, {Sasselov}, {Sozzetti},
  {Szentgyorgyi}, {Udry}, \& {Walsworth}}]{Collier-Cameron2019aa}
{Collier Cameron}, A., {Mortier}, A., {Phillips}, D., {et~al.} 2019, \mnras,
  487, 1082

\bibitem[{{Cretignier} {et~al.}(2020){Cretignier}, {Dumusque}, {Allart},
  {Pepe}, \& {Lovis}}]{Cretignier:2020aa}
{Cretignier}, M., {Dumusque}, X., {Allart}, R., {Pepe}, F., \& {Lovis}, C.
  2020, \aap, 633, A76

\bibitem[{{de Beurs} {et~al.}(2020){de Beurs}, {Vanderburg}, {Shallue},
  {Dumusque}, {Collier Cameron}, {Buchhave}, {Cosentino}, {Ghedina}, {Haywood},
  {Langellier}, {Latham}, {L{\'o}pez-Morales}, {Mayor}, {Micela}, {Milbourne},
  {Mortier}, {Molinari}, {Pepe}, {Phillips}, {Pinamonti}, {Piotto}, {Rice},
  {Sasselov}, {Sozzetti}, {Udry}, \& {Watson}}]{Beurs:2020aa}
{de Beurs}, Z.~L., {Vanderburg}, A., {Shallue}, C.~J., {et~al.} 2020, arXiv
  e-prints, arXiv:2011.00003

\bibitem[{{Delisle} {et~al.}(2020){Delisle}, {Hara}, \&
  {S{\'e}gransan}}]{Delisle:2020ab}
{Delisle}, J.~B., {Hara}, N., \& {S{\'e}gransan}, D. 2020, \aap, 638, A95

\bibitem[{{Delisle} {et~al.}(2018){Delisle}, {S{\'e}gransan}, {Dumusque},
  {Diaz}, {Bouchy}, {Lovis}, {Pepe}, {Udry}, {Alonso}, {Benz}, {Coffinet},
  {Cameron}, {Deleuil}, {Figueira}, {Gillon}, {Curto}, {Mayor}, {Mordasini},
  {Motalebi}, {Moutou}, {Pollacco}, {Pompei}, {Queloz}, {Santos}, \&
  {Wyttenbach}}]{Delisle:2018aa}
{Delisle}, J.-B., {S{\'e}gransan}, D., {Dumusque}, X., {et~al.} 2018, \aap,
  614, A133

\bibitem[{{Desort} {et~al.}(2007){Desort}, {Lagrange}, {Galland}, {Udry}, \&
  {Mayor}}]{Desort-2007}
{Desort}, M., {Lagrange}, A.-M., {Galland}, F., {Udry}, S., \& {Mayor}, M.
  2007, \aap, 473, 983

\bibitem[{{D{\'{\i}}az} {et~al.}(2016){D{\'{\i}}az}, {S{\'e}gransan}, {Udry},
  {Lovis}, {Pepe}, {Dumusque}, {Marmier}, {Alonso}, {Benz}, {Bouchy},
  {Coffinet}, {Collier Cameron}, {Deleuil}, {Figueira}, {Gillon}, {Lo Curto},
  {Mayor}, {Mordasini}, {Motalebi}, {Moutou}, {Pollacco}, {Pompei}, {Queloz},
  {Santos}, \& {Wyttenbach}}]{Diaz:2016ab}
{D{\'{\i}}az}, R.~F., {S{\'e}gransan}, D., {Udry}, S., {et~al.} 2016, \aap,
  585, A134

\bibitem[{{Dumusque}(2018)}]{Dumusque:2018aa}
{Dumusque}, X. 2018, \aap, 620, A47

\bibitem[{{Dumusque} {et~al.}(2015){Dumusque}, {Glenday}, {Phillips},
  {Buchschacher}, {Collier Cameron}, {Cecconi}, {Charbonneau}, {Cosentino},
  {Ghedina}, {Latham}, {Li}, {Lodi}, {Lovis}, {Molinari}, {Pepe}, {Udry},
  {Sasselov}, {Szentgyorgyi}, \& {Walsworth}}]{Dumusque-2015b}
{Dumusque}, X., {Glenday}, A., {Phillips}, D.~F., {et~al.} 2015, \apjl, 814,
  L21

\bibitem[{{Dumusque} {et~al.}(2011{\natexlab{a}}){Dumusque}, {Lovis},
  {S{\'e}gransan}, {Mayor}, {Udry}, {Benz}, {Bouchy}, {Lo Curto}, {Mordasini},
  {Pepe}, {Queloz}, {Santos}, \& {Naef}}]{Dumusque-2011c}
{Dumusque}, X., {Lovis}, C., {S{\'e}gransan}, D., {et~al.} 2011{\natexlab{a}},
  \aap, 535, A55

\bibitem[{{Dumusque} {et~al.}(2012){Dumusque}, {Pepe}, {Lovis}, {Segransan},
  {Sahlmann}, {Benz}, {Bouchy}, {Mayor}, {Queloz}, {Santos}, \&
  {Udry}}]{Dumusque-2012}
{Dumusque}, X., {Pepe}, F., {Lovis}, C., {et~al.} 2012, \nat, 491, 207

\bibitem[{{Dumusque} {et~al.}(2011{\natexlab{b}}){Dumusque}, {Udry}, {Lovis},
  {Santos}, \& {Monteiro}}]{Dumusque-2011a}
{Dumusque}, X., {Udry}, S., {Lovis}, C., {Santos}, N.~C., \& {Monteiro},
  M.~J.~P.~F.~G. 2011{\natexlab{b}}, \aap, 525, A140

\bibitem[{{Feng} {et~al.}(2017){Feng}, {Tuomi}, {Jones}, {Barnes},
  {Anglada-Escud{\'e}}, {Vogt}, \& {Butler}}]{Feng:2017ac}
{Feng}, F., {Tuomi}, M., {Jones}, H.~R.~A., {et~al.} 2017, \aj, 154, 135

\bibitem[{{Fischer} {et~al.}(2016){Fischer}, {Anglada-Escude}, {Arriagada},
  {Baluev}, {Bean}, {Bouchy}, {Buchhave}, {Carroll}, {Chakraborty}, {Crepp},
  {Dawson}, {Diddams}, {Dumusque}, {Eastman}, {Endl}, {Figueira}, {Ford},
  {Foreman-Mackey}, {Fournier}, {Furesz}, {Gaudi}, {Gregory}, {Grundahl},
  {Hatzes}, {Hebrard}, {Herrero}, {Hogg}, {Howard}, {Johnson}, {Jorden},
  {Jurgenson}, {Latham}, {Laughlin}, {Loredo}, {Lovis}, {Mahadevan},
  {McCracken}, {Pepe}, {Perez}, {Phillips}, {Plavchan}, {Prato}, {Quirrenbach},
  {Reiners}, {Robertson}, {Santos}, {Sawyer}, {Segransan}, {Sozzetti},
  {Steinmetz}, {Szentgyorgyi}, {Udry}, {Valenti}, {Wang}, {Wittenmyer}, \&
  {Wright}}]{Fischer-2016}
{Fischer}, D., {Anglada-Escude}, G., {Arriagada}, P., {et~al.} 2016, ArXiv
  e-prints [\eprint[arXiv]{1602.07939}]

\bibitem[{{Foreman-Mackey} {et~al.}(2014){Foreman-Mackey}, {Hogg}, \&
  {Morton}}]{Foreman-Mackey-2014aa}
{Foreman-Mackey}, D., {Hogg}, D.~W., \& {Morton}, T.~D. 2014, \apj, 795, 64

\bibitem[{{Haywood} {et~al.}(2016){Haywood}, {Collier Cameron}, {Unruh},
  {Lovis}, {Lanza}, {Llama}, {Deleuil}, {Fares}, {Gillon}, {Moutou}, {Pepe},
  {Pollacco}, {Queloz}, \& {S{\'e}gransan}}]{Haywood-2016}
{Haywood}, R.~D., {Collier Cameron}, A., {Unruh}, Y.~C., {et~al.} 2016, \mnras,
  457, 3637

\bibitem[{{Haywood} {et~al.}(2020){Haywood}, {Milbourne}, {Saar}, {Mortier},
  {Phillips}, {Charbonneau}, {Collier Cameron}, {Cegla}, {Meunier}, \&
  {Palumbo}}]{Haywood:2020aa}
{Haywood}, R.~D., {Milbourne}, T.~W., {Saar}, S.~H., {et~al.} 2020, arXiv
  e-prints, arXiv:2005.13386

\bibitem[{{Hogg} {et~al.}(2010){Hogg}, {Bovy}, \& {Lang}}]{Hogg:2010aa}
{Hogg}, D.~W., {Bovy}, J., \& {Lang}, D. 2010, arXiv e-prints, arXiv:1008.4686

\bibitem[{{Langellier} {et~al.}(2020){Langellier}, {Milbourne}, {Phillips},
  {Haywood}, {Saar}, {Mortier}, {Malavolta}, {Thompson}, {Collier Cameron},
  {Dumusque}, {Cegla}, {Latham}, {Maldonado}, {Watson}, {Cecconi},
  {Charbonneau}, {Cosentino}, {Ghedina}, {Gonzalez}, {Li}, {Lodi},
  {L{\'o}pez-Morales}, {Micela}, {Molinari}, {Pepe}, {Poretti}, {Rice},
  {Sasselov}, {Sozzetti}, {Udry}, \& {Walsworth}}]{Langellier:2020aa}
{Langellier}, N., {Milbourne}, T.~W., {Phillips}, D.~F., {et~al.} 2020, arXiv
  e-prints, arXiv:2008.05970

\bibitem[{{Lean} \& {Brueckner}(1989)}]{Lean:1989aa}
{Lean}, J.~L. \& {Brueckner}, G.~E. 1989, \apj, 337, 568

\bibitem[{{Lefebvre} {et~al.}(2008){Lefebvre}, {Garc{\'{\i}}a},
  {Jim{\'e}nez-Reyes}, {Turck-Chi{\`e}ze}, \& {Mathur}}]{Lefebvre-2008}
{Lefebvre}, S., {Garc{\'{\i}}a}, R.~A., {Jim{\'e}nez-Reyes}, S.~J.,
  {Turck-Chi{\`e}ze}, S., \& {Mathur}, S. 2008, \aap, 490, 1143

\bibitem[{{Lindegren} \& {Dravins}(2003)}]{Lindegren-2003}
{Lindegren}, L. \& {Dravins}, D. 2003, \aap, 401, 1185

\bibitem[{{Lockwood} {et~al.}(2007){Lockwood}, {Skiff}, {Henry}, {Henry},
  {Radick}, {Baliunas}, {Donahue}, \& {Soon}}]{Lockwood-2007}
{Lockwood}, G.~W., {Skiff}, B.~A., {Henry}, G.~W., {et~al.} 2007, \apjs, 171,
  260

\bibitem[{{Lovis} {et~al.}(2011){Lovis}, {Dumusque}, {Santos}, {Bouchy},
  {Mayor}, {Pepe}, {Queloz}, {S{\'e}gransan}, \& {Udry}}]{Lovis-2011b}
{Lovis}, C., {Dumusque}, X., {Santos}, N.~C., {et~al.} 2011, ArXiv e-prints
  [\eprint[arXiv]{1107.5325}]

\bibitem[{{Lovis} \& {Fischer}(2010)}]{Lovis:2010aa}
{Lovis}, C. \& {Fischer}, D. 2010, {Radial Velocity Techniques for Exoplanets},
  ed. S.~{Seager}, 27--53

\bibitem[{{Lovis} \& {Pepe}(2007)}]{Lovis-2007b}
{Lovis}, C. \& {Pepe}, F. 2007, \aap, 468, 1115

\bibitem[{{Maldonado} {et~al.}(2019){Maldonado}, {Phillips}, {Dumusque},
  {Collier Cameron}, {Haywood}, {Lanza}, {Micela}, {Mortier}, {Saar},
  {Sozzetti}, {Rice}, {Milbourne}, {Cecconi}, {Cegla}, {Cosentino}, {Costes},
  {Ghedina}, {Gonzalez}, {Guerra}, {Hern{\'a}ndez}, {Li}, {Lodi}, {Malavolta},
  {Molinari}, {Pepe}, {Piotto}, {Poretti}, {Sasselov}, {San Juan}, {Thompson},
  {Udry}, \& {Watson}}]{Maldonado:2019aa}
{Maldonado}, J., {Phillips}, D.~F., {Dumusque}, X., {et~al.} 2019, \aap, 627,
  A118

\bibitem[{{Meunier} {et~al.}(2010{\natexlab{a}}){Meunier}, {Desort}, \&
  {Lagrange}}]{Meunier-2010a}
{Meunier}, N., {Desort}, M., \& {Lagrange}, A.-M. 2010{\natexlab{a}}, \aap,
  512, A39

\bibitem[{{Meunier} {et~al.}(2010{\natexlab{b}}){Meunier}, {Lagrange}, \&
  {Desort}}]{Meunier-2010b}
{Meunier}, N., {Lagrange}, A., \& {Desort}, M. 2010{\natexlab{b}}, \aap, 519,
  A66+

\bibitem[{{Meunier} \& {Lagrange}(2020)}]{Meunier:2020aa}
{Meunier}, N. \& {Lagrange}, A.~M. 2020, \aap, 638, A54

\bibitem[{{Meunier} {et~al.}(2017{\natexlab{a}}){Meunier}, {Lagrange}, \&
  {Borgniet}}]{Meunier:2017aa}
{Meunier}, N., {Lagrange}, A.-M., \& {Borgniet}, S. 2017{\natexlab{a}}, \aap,
  607, A6

\bibitem[{{Meunier} {et~al.}(2015){Meunier}, {Lagrange}, {Borgniet}, \&
  {Rieutord}}]{Meunier-2015}
{Meunier}, N., {Lagrange}, A.-M., {Borgniet}, S., \& {Rieutord}, M. 2015, \aap,
  583, A118

\bibitem[{{Meunier} {et~al.}(2017{\natexlab{b}}){Meunier}, {Lagrange}, {Mbemba
  Kabuiku}, {Alex}, {Mignon}, \& {Borgniet}}]{Meunier:2017ab}
{Meunier}, N., {Lagrange}, A.-M., {Mbemba Kabuiku}, L., {et~al.}
  2017{\natexlab{b}}, \aap, 597, A52

\bibitem[{{Miklos} {et~al.}(2020){Miklos}, {Milbourne}, {Haywood}, {Phillips},
  {Saar}, {Meunier}, {Cegla}, {Dumusque}, {Langellier}, {Maldonado},
  {Malavolta}, {Mortier}, {Thompson}, {Watson}, {Cecconi}, {Cosentino},
  {Ghedina}, {Li}, {L{\'o}pez-Morales}, {Molinari}, {Poretti}, {Sasselov},
  {Sozzetti}, \& {Walsworth}}]{Miklos:2020aa}
{Miklos}, M., {Milbourne}, T.~W., {Haywood}, R.~D., {et~al.} 2020, \apj, 888,
  117

\bibitem[{{Milbourne} {et~al.}(2019){Milbourne}, {Haywood}, {Phillips}, {Saar},
  {Cegla}, {Cameron}, {Costes}, {Dumusque}, {Langellier}, {Latham},
  {Maldonado}, {Malavolta}, {Mortier}, {Palumbo}, {Thompson}, {Watson},
  {Bouchy}, {Buchschacher}, {Cecconi}, {Charbonneau}, {Cosentino}, {Ghedina},
  {Glenday}, {Gonzalez}, {Li}, {Lodi}, {L{\'o}pez-Morales}, {Lovis}, {Mayor},
  {Micela}, {Molinari}, {Pepe}, {Piotto}, {Rice}, {Sasselov}, {S{\'e}gransan},
  {Sozzetti}, {Szentgyorgyi}, {Udry}, \& {Walsworth}}]{Milbourne:2019aa}
{Milbourne}, T.~W., {Haywood}, R.~D., {Phillips}, D.~F., {et~al.} 2019, \apj,
  874, 107

\bibitem[{National Academies~of Sciences(2018)}]{nas2018}
National Academies~of Sciences, Engineering, M. 2018, in Exoplanet science
  strategy (Consensus study report), Washington, DC: The National Academies
  Press.

\bibitem[{{Noyes} {et~al.}(1984){Noyes}, {Hartmann}, {Baliunas}, {Duncan}, \&
  {Vaughan}}]{Noyes-1984}
{Noyes}, R.~W., {Hartmann}, L.~W., {Baliunas}, S.~L., {Duncan}, D.~K., \&
  {Vaughan}, A.~H. 1984, \apj, 279, 763

\bibitem[{{Palmer} \& {Engleman}(1983)}]{Palmer:1983aa}
{Palmer}, B.~A. \& {Engleman}, R. 1983, {Atlas of the Thorium spectrum}

\bibitem[{{Pepe} {et~al.}(2020){Pepe}, {Cristiani}, {Rebolo}, {Santos},
  {Dekker}, {Cabral}, {Di Marcantonio}, {Figueira}, {Lo Curto}, {Lovis},
  {Mayor}, {M{\'e}gevand}, {Molaro}, {Riva}, {Zapatero Osorio}, {Amate},
  {Manescau}, {Pasquini}, {Zerbi}, {Adibekyan}, {Abreu}, {Affolter}, {Alibert},
  {Aliverti}, {Allart}, {Allende Prieto}, {{\'A}lvarez}, {Alves}, {Avila},
  {Baldini}, {Bandy}, {Barros}, {Benz}, {Bianco}, {Borsa}, {Bourrier},
  {Bouchy}, {Broeg}, {Calderone}, {Cirami}, {Coelho}, {Conconi}, {Coretti},
  {Cumani}, {Cupani}, {D'Odorico}, {Damasso}, {Deiries}, {Delabre},
  {Demangeon}, {Dumusque}, {Ehrenreich}, {Faria}, {Fragoso}, {Genolet},
  {Genoni}, {G{\'e}nova Santos}, {Gonz{\'a}lez Hern{\'a}ndez}, {Hughes},
  {Iwert}, {Kerber}, {Knudstrup}, {Landoni}, {Lavie}, {Lillo-Box}, {Lizon},
  {Maire}, {Martins}, {Mehner}, {Micela}, {Modigliani}, {Monteiro}, {Monteiro},
  {Moschetti}, {Murphy}, {Nunes}, {Oggioni}, {Oliveira}, {Oshagh}, {Pall{\'e}},
  {Pariani}, {Poretti}, {Rasilla}, {Rebord{\~a}o}, {Redaelli}, {Santana
  Tschudi}, {Santin}, {Santos}, {S{\'e}gransan}, {Schmidt}, {Segovia},
  {Sosnowska}, {Sozzetti}, {Sousa}, {Span{\`o}}, {Su{\'a}rez Mascare{\~n}o},
  {Tabernero}, {Tenegi}, {Udry}, \& {Zanutta}}]{Pepe:2020aa}
{Pepe}, F., {Cristiani}, S., {Rebolo}, R., {et~al.} 2020, arXiv e-prints,
  arXiv:2010.00316

\bibitem[{Pepe {et~al.}(2002)Pepe, Mayor, Galland, Naef, Queloz, Santos, Udry,
  \& Burnet}]{Pepe-2002a}
Pepe, F., Mayor, M., Galland, F., {et~al.} 2002, \aap, 388, 632

\bibitem[{{Phillips} {et~al.}(2016){Phillips}, {Glenday}, {Dumusque},
  {Buchschacher}, {Collier Cameron}, {Cecconi}, {Charbonneau}, {Cosentino},
  {Ghedina}, {Haywood}, {Latham}, {Li}, {Lodi}, {Lovis}, {Molinari}, {Pepe},
  {Sasselov}, {Szentgyorgyi}, {Udry}, \& {Walsworth}}]{Phillips:2016aa}
{Phillips}, D.~F., {Glenday}, A.~G., {Dumusque}, X., {et~al.} 2016, in Society
  of Photo-Optical Instrumentation Engineers (SPIE) Conference Series, Vol.
  9912, Advances in Optical and Mechanical Technologies for Telescopes and
  Instrumentation II, 99126Z

\bibitem[{{Plavchan} {et~al.}(2018){Plavchan}, {Cale}, {Newman}, {Hamze},
  {Latouf}, {Matzko}, {Beichman}, {Ciardi}, {Purcell}, {Lightsey}, {Cegla},
  {Dumusque}, {Bourrier}, {Dressing}, {Gao}, {Vasisht}, {Leifer}, {Wang},
  {Gagne}, {Thompson}, {Crass}, {Bechter}, {Bechter}, {Blake}, {Halverson},
  {Mayo}, {Beatty}, {Wright}, {Wise}, {Tanner}, {Eastman}, {Quinn}, {Fischer},
  {Basu}, {Sanchez-Maes}, {Howard}, {Vahala}, {Wang}, {Diddams}, {Papp},
  {Pope}, {Martin}, \& {Murphy}}]{Plavchan:2018aa}
{Plavchan}, P., {Cale}, B., {Newman}, P., {et~al.} 2018, arXiv e-prints,
  arXiv:1803.03960

\bibitem[{{Press} \& {Rybicki}(1989)}]{Press-1989}
{Press}, W.~H. \& {Rybicki}, G.~B. 1989, \apj, 338, 277

\bibitem[{{Rajpaul} {et~al.}(2015){Rajpaul}, {Aigrain}, {Osborne}, {Reece}, \&
  {Roberts}}]{Rajpaul-2015}
{Rajpaul}, V., {Aigrain}, S., {Osborne}, M.~A., {Reece}, S., \& {Roberts}, S.
  2015, \mnras, 452, 2269

\bibitem[{{Redman} {et~al.}(2014){Redman}, {Nave}, \&
  {Sansonetti}}]{Redman:2014aa}
{Redman}, S.~L., {Nave}, G., \& {Sansonetti}, C.~J. 2014, \apjs, 211, 4

\bibitem[{{Reiners} {et~al.}(2010){Reiners}, {Bean}, {Huber}, {Dreizler},
  {Seifahrt}, \& {Czesla}}]{Reiners-2010}
{Reiners}, A., {Bean}, J.~L., {Huber}, K.~F., {et~al.} 2010, \apj, 710, 432

\bibitem[{{Rieger} {et~al.}(1984){Rieger}, {Share}, {Forrest}, {Kanbach},
  {Reppin}, \& {Chupp}}]{Rieger:1984aa}
{Rieger}, E., {Share}, G.~H., {Forrest}, D.~J., {et~al.} 1984, \nat, 312, 623

\bibitem[{{Saar} \& {Donahue}(1997)}]{Saar-1997b}
{Saar}, S.~H. \& {Donahue}, R.~A. 1997, \apj, 485, 319

\bibitem[{{Strassmeier} {et~al.}(2015){Strassmeier}, {Ilyin}, {J{\"a}rvinen},
  {Weber}, {Woche}, {Barnes}, {Bauer}, {Beckert}, {Bittner}, {Bredthauer},
  {Carroll}, {Denker}, {Dionies}, {DiVarano}, {D{\"o}scher}, {Fechner},
  {Feuerstein}, {Granzer}, {Hahn}, {Harnisch}, {Hofmann}, {Lesser}, {Paschke},
  {Pankratow}, {Plank}, {Pl{\"u}schke}, {Popow}, {Sablowski}, \&
  {Storm}}]{Strassmeier-2015}
{Strassmeier}, K.~G., {Ilyin}, I., {J{\"a}rvinen}, A., {et~al.} 2015, ArXiv
  e-prints [\eprint[arXiv]{1505.06492}]

\bibitem[{{Vaughan} {et~al.}(1978){Vaughan}, {Preston}, \&
  {Wilson}}]{Vaughan-1978}
{Vaughan}, A.~H., {Preston}, G.~W., \& {Wilson}, O.~C. 1978, \pasp, 90, 267

\bibitem[{{Wilson}(1978)}]{Wilson-1978}
{Wilson}, O.~C. 1978, \apj, 226, 379

\bibitem[{{Zechmeister} \& {K{\"u}rster}(2009)}]{Zechmeister-2009}
{Zechmeister}, M. \& {K{\"u}rster}, M. 2009, \aap, 496, 577

\bibitem[{{Zechmeister} {et~al.}(2018){Zechmeister}, {Reiners}, {Amado},
  {Azzaro}, {Bauer}, {B{\'e}jar}, {Caballero}, {Guenther}, {Hagen}, {Jeffers},
  {Kaminski}, {K{\"u}rster}, {Launhardt}, {Montes}, {Morales}, {Quirrenbach},
  {Reffert}, {Ribas}, {Seifert}, {Tal-Or}, \& {Wolthoff}}]{Zechmeister:2018aa}
{Zechmeister}, M., {Reiners}, A., {Amado}, P.~J., {et~al.} 2018, \aap, 609, A12

\end{thebibliography}

\begin{appendix}

\section{Description of the data released}\label{app:data_release}

The data release consists of the following data products:
\begin{itemize}
\item The extracted echelle-order spectra, corrected from the instrumental blaze, in the Earth rest-frame. These products are called S2D spectra due to their two dimensional shape.
The first and second extensions in the FITS file contain the blaze-corrected extracted flux per pixel and corresponding error for each spectral order. 
The error corresponds to the photon plus read-out noises of the detector added in quadrature, and divided by the blaze, so that the corresponding error can be directly used with the flux given in the first extension of the FITS file. The third extension corresponds to the quality flag of the pixels for each order, zero being good, and anything else being bad. Hot and bad pixels are flagged that way. Extensions four and five are the wavelength solution in the vacuum and in air, and extensions six and seven are the width of pixels in wavelength in the vacuum and in air, respectively. We note that all wavelengths are in \r{A}ngstr{\"o}m. Because of dispersion, the size of each pixel in \r{A}ngstr{\"o}m will change with wavelength, which implies that for a given order, the continuum of an S2D spectrum will show a significant slope. To get a flat continuum, the easiest way is to transform the wavelengths in a logarithmic scale.

\item The extracted merged spectra, corrected from the instrumental blaze, in the Solar System barycentric rest-frame. These products are called S1D spectra. The only extension in the FITS file includes the wavelength in the vacuum and in air for each point of the merged spectrum, its flux and the quality flag of the point, as defined in the first item above. We note that merged-1d spectra are interpolated on a grid constant in velocity space and not in wavelength space. The step between each point is 0.82\kms, equivalent to the width of a pixel in velocity space.
\item The CCFs obtained by cross-correlating the S2D spectra with a synthetic mask derived from solar observations, thus optimized for the Sun. The first extension in the FITS file gives the CCF measured for each echelle order, with a step of 0.82\kms, in 
addition to the photon-noise weighted average CCF over all orders. Therefore, the extension has the shape $N_{CCF} \times (N_{ord}+1)$, where $N_{CCF}$ is the number of points of the CCF and $N_{ord}$ is the number of echelle orders, 69 for HARPS-N. The second extension gives the photon noise plus read-out errors, and the third one gives the quality flag of each point as defined in the first item above.
\end{itemize}
In addition to those data products, we also release the time-series of relevant quantities such as the RV, the BIS SPAN, the FWHM, different activity indicators. They are all listed in Table~\ref{tab:temeseries_list} with their descriptions.

The data can be accessed using the Data Analysis Center for Exoplanet (DACE) web-platform at this address \url{https://dace.unige.ch/sun/}, or using the DACE python API (\url{https://dace.unige.ch/pythonAPI/} with the solar spectroscopy module). Bulk downloads are available on the main web page, to download all the available products (S2D, S1D and CCF), only the CCF products, or the time-series. Using the solar spectroscopy data base (go to the main web page and then follow the link to the \emph{Solar spectroscopy database}), it is possible to scan through all the released data, select specific observations by using filters (magnifying glass on the right of each column) and download the S2D, S1D or CCF products of specific observations by selecting the rows of interest (click on them), and then using one of the S2D, S1D or CCF buttons on the top right corner of the table. We note that although by default only eight columns are shown on the \emph{Solar spectroscopy database} web page, it is possible to get access to more information by clicking on the gearwheel on the top left corner of the table. All this can also be done using the DACE python API, for people that want to directly access to the data within a python script (\url{https://dace.unige.ch/pythonAPI/?tutorialId=22}). A small tutorial to perform basic analysis on the solar data released is also available at this address \url{https://dace-preview.unige.ch/pythonAPI/?tutorialId=23}.
\begin{table*}
	\footnotesize
	\caption{Description of the time series available for the HARPS-N solar data.}            
	\label{tab:temeseries_list}    
	\centering                         
	\begin{tabular}{p{2.5cm}p{11cm}p{1.5cm}p{1.5cm}}
		\hline\hline
		Name & Description & Unit & Type \\
		\hline
        \emph{Filename} & 
        Name of the observation
        & - & STRING \\ 
        \emph{Date BJD} & 
        Barycentric Julian Date minus 2400000
        & days & FLOAT \\ 
        \emph{RV raw} & 
        RV of the Sun in the Solar System barycentric rest frame
        & \ms & FLOAT \\ 
        \emph{RV} & 
        RV of the Sun in the heliocentric rest frame, corrected for differential extinction. This value is obtained by substracting \emph{Berv bary to helio} and \emph{RV diff extinction} to \emph{RV raw}
        & \ms & FLOAT \\ 
        \emph{RV err} & 
        RV error 
        & \ms & FLOAT \\ 
        \emph{Rhk} & 
        \logrhk calcium activity index
        & dex & FLOAT \\      
        \emph{Rhk err} & 
        \logrhk calcium activity index error
        & dex & FLOAT \\ 
        \emph{Smw} & 
        S Mount Wilson calcium activity index
        & - & FLOAT \\      
        \emph{Smw err} & 
        S Mount Wilson calcium activity index error
        & - & FLOAT \\ 
        \emph{Bis Span} & 
        Bisector span of the CCF
        & m/s & FLOAT \\ 
        \emph{Bis Span err} & 
        Bisector span error of the CCF
        & m/s & FLOAT \\ 
        \emph{FWHM raw} & 
        Raw FWHM of the CCF
        & m/s & FLOAT \\ 
        \emph{FWHM} & 
        FWHM of the CCF, corrected for the solar ecliptic obliquity and Earth orbit eccentricity. See section 3.2 in Collier-Cameron et al. (2019) for more information. We note that in that paper, the authors have to optimize the value of $\gamma$ corresponding to the fraction of the observed \vsini. They found a value of 1.04 for the old DRS solar data. Performing the same optimisation for the new DRS data lead to a value of 1.15.
        & m/s & FLOAT \\ 
        \emph{FWHM err} & 
        FWHM error of the CCF
        & m/s & FLOAT \\ 
        \emph{Contrast raw} & 
        Raw contrast of the CCF
        & \% & FLOAT \\ 
        \emph{Contrast} & 
        Contrast of the CCF, corrected for the FWHM correction so that the equivalent width of the CCF is conserved. See section 3.2 in Collier-Cameron et al. (2019) for more information
        & \% & FLOAT \\ 
        \emph{Contrast err} & 
        Contrast error of the CCF
        & \% & FLOAT \\ 
        \emph{Berv} & 
        Barycentric Earth RV correction
        & m/s & FLOAT \\
        \emph{Berv bary to helio} & 
        Correction to change from the Solar System barycentric to heliocentric rest frame. To change from the heliocentric to barycentric rest frame, just add this term to RV
        & m/s & FLOAT \\
        \emph{RV diff extinction} & 
        Estimation of the RV effect induced by differential extinction. See section 2.4  in Collier-Cameron et al. (2019) for more information. To include the effect of differential extinction, just add this value to \emph{RV}
        & m/s & FLOAT \\
        \emph{Airmass} & 
        Sun airmass
        & - & FLOAT \\
        \emph{Coordinates} & 
        Sun coordinates
        & hms / dms & STRING \\
        \emph{Texp} & 
        Exposure time
        & sec & FLOAT \\
        \emph{Sn order 10} & 
        Signal to noise in order 10
        & - & FLOAT \\
        \emph{Sn order 20} & 
        Signal to noise in order 20
        & - & FLOAT \\
        \emph{Sn order 30} & 
        Signal to noise in order 30
        & - & FLOAT \\
        \emph{Sn order 40} & 
        Signal to noise in order 40
        & - & FLOAT \\
        \emph{Sn order 50} & 
        Signal to noise in order 50
        & - & FLOAT \\
        \emph{Sn order 60} & 
        Signal to noise in order 60
        & - & FLOAT \\
        \emph{Obs quality} & 
        Quality flag to assess the observation's quality (no clouds, no calima). Data with values >= 0.99 are all excellent. See section 2.3 in Collier-Cameron et al. (2019) for more information.
        & - & FLOAT \\
        \emph{DRS quality} & 
        Quality flag of the data reduction software (true for good, false for bad)
        & - & BOOLEAN \\

		\hline
	\end{tabular}
\end{table*}
%

\section{Wavelength solution derived using the drift map method}\label{app:drift_map_method}

\begin{figure*}[!h]
        \center
	\includegraphics[width=14.2cm]{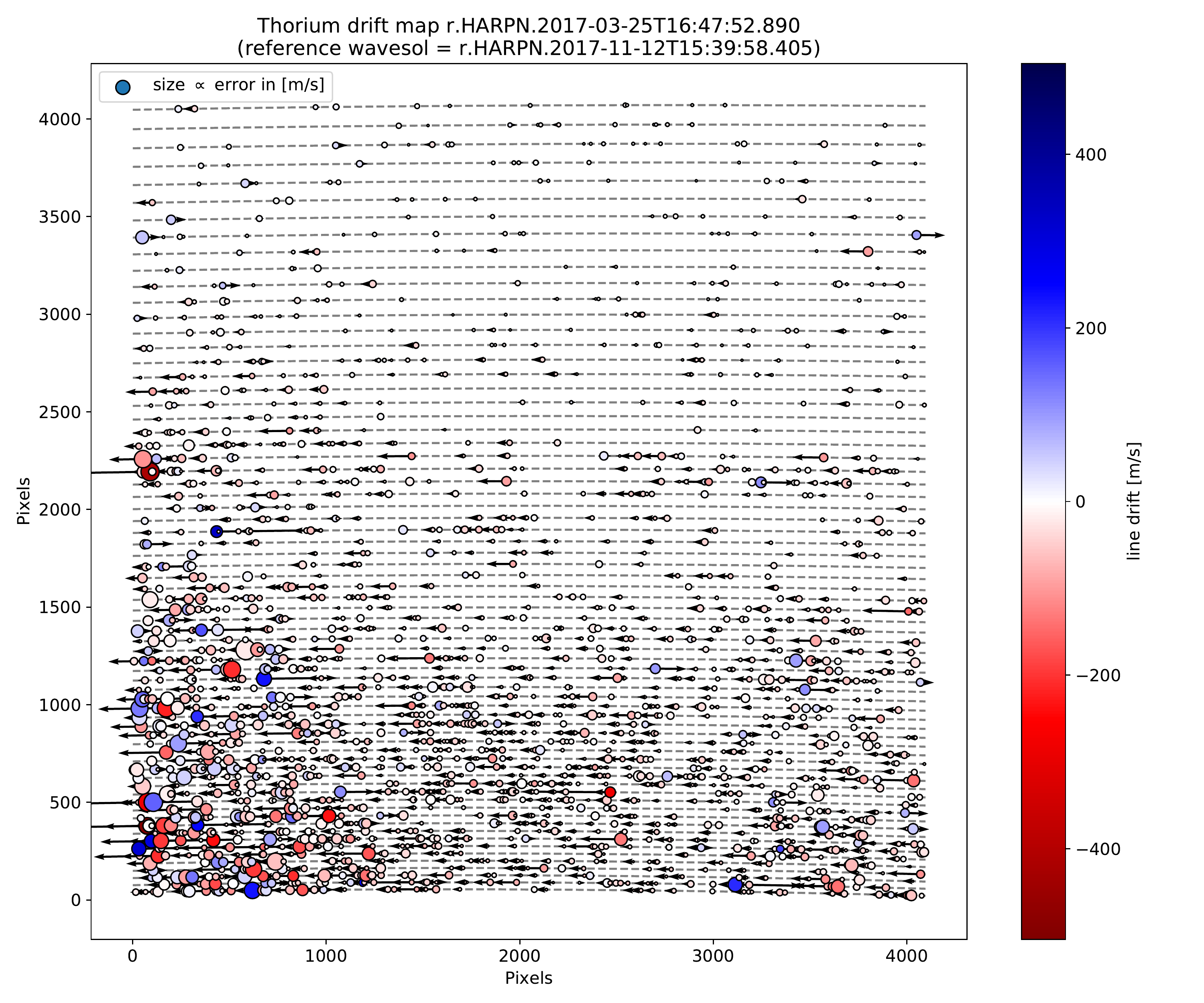}
	\includegraphics[width=9cm]{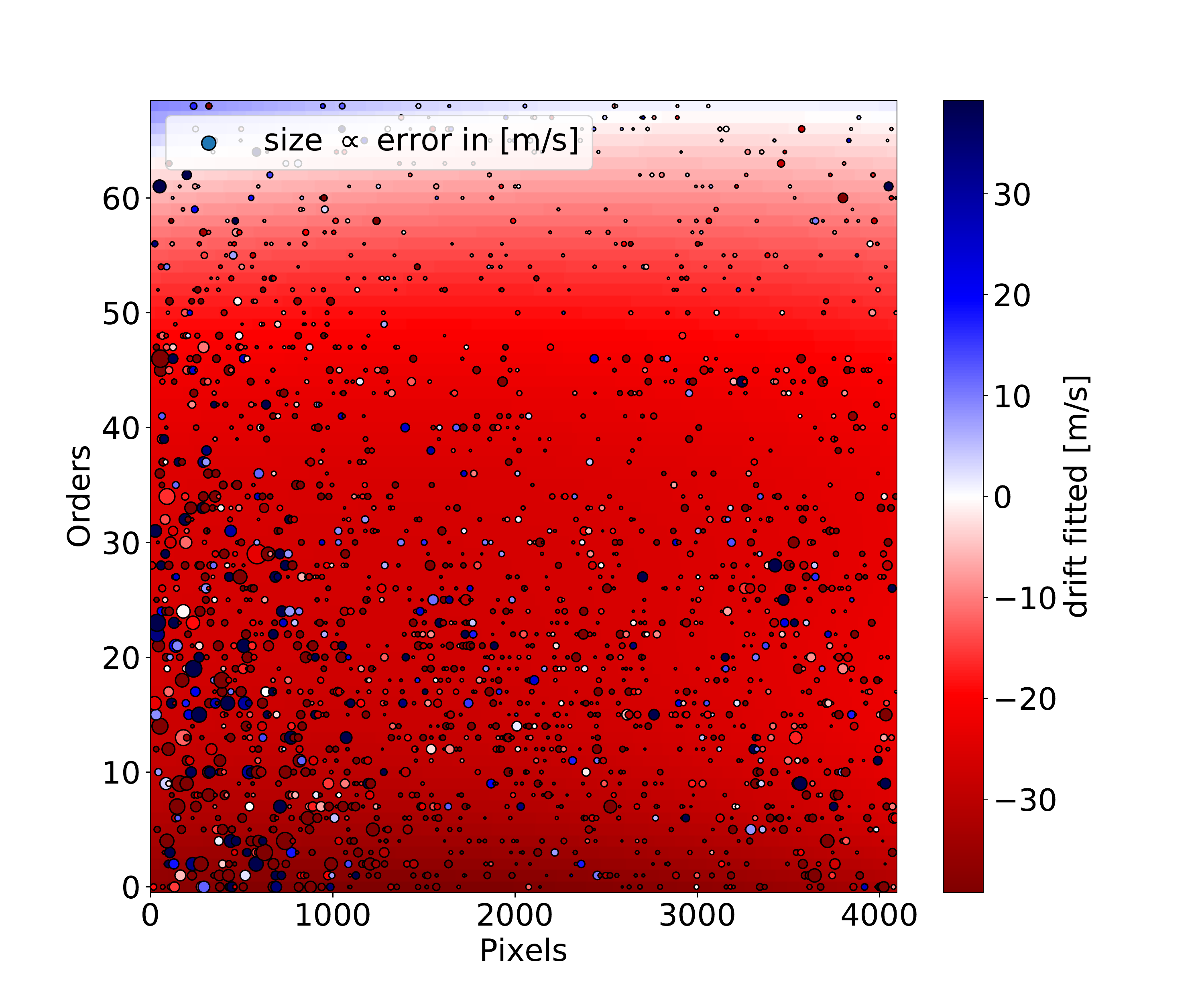}
	\includegraphics[width=9cm]{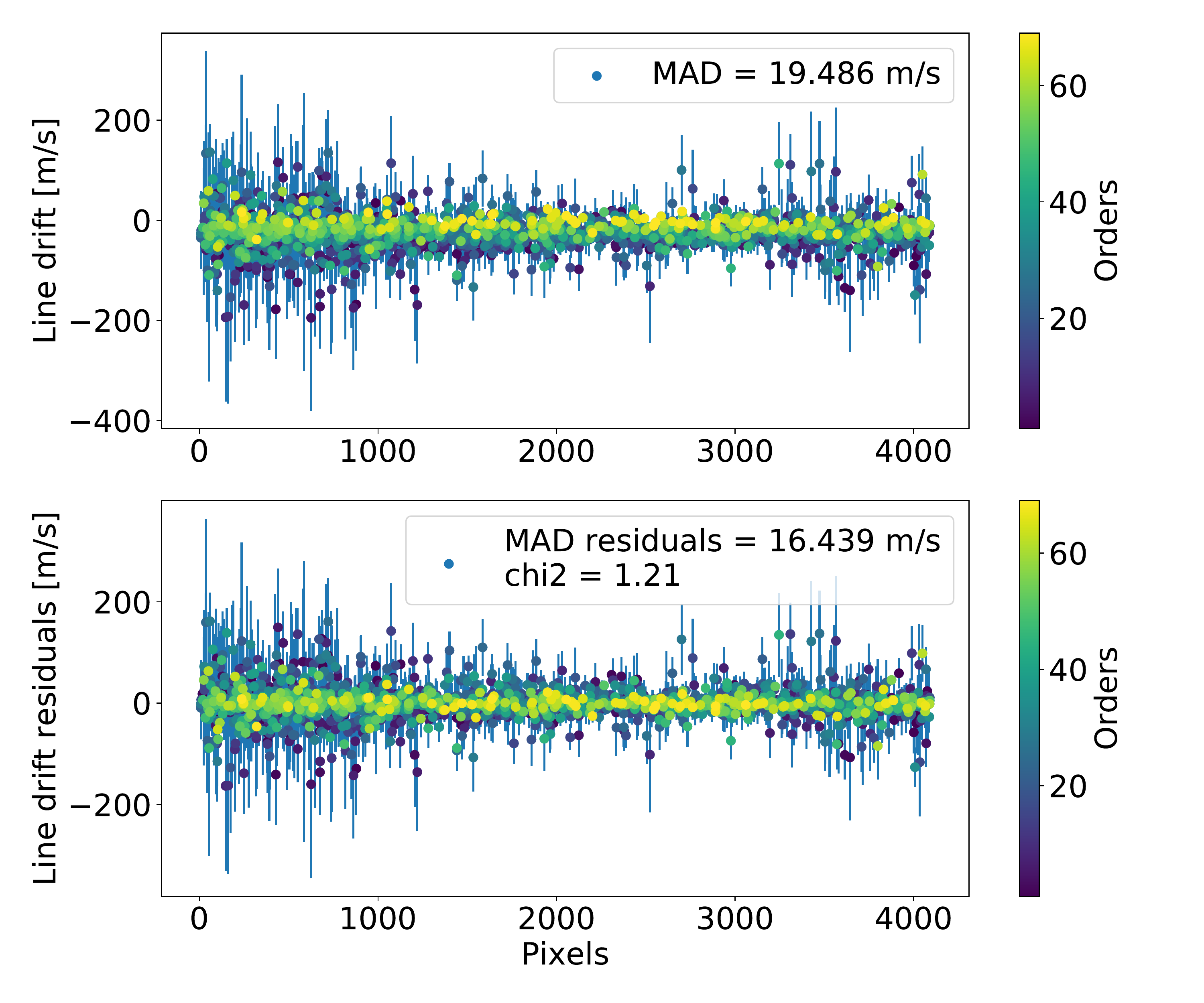}
	\caption{Derivation of the wavelength solution for the thorium calibration HARPN.2017-03-25T16-47-52.890.fits using the drift map method with the frame HARPN.2017-11-12T15:39:58.405.fits selected as reference. \emph{Top:} First, we measure the drift of all the thorium lines between those two calibrations. Each thorium line is represented as a dot with size proportional to error in drift and colour to the drift measured. We note that the thorium lines have been represented at their exact location on the detector using the localisation of the orders (grey dotted lines). \emph{Bottom left:} Then, we fit a smooth two-dimensional polynomial on the measured drift of all thorium lines. The corresponding wavelength solution is obtained by shifting the reference wavelength solution with the fitted two dimensional polynomial. \emph{Bottom right:} Measured drift on the thorium lines between the thorium calibration and the reference before adjusting the two dimensional polynomial (top) and after fitting for it (bottom). As we see, the $\chi^2$ of the two dimensional polynomial fit is close to one, and the residuals show a scatter, measure with the MAD, which is significantly reduced.}
	\label{fig:drift_map_method}
\end{figure*}

In Fig.~\ref{fig:drift_map_method} we show the different steps performed to derive and validate the quality of a wavelength solution using the drift map method. The first step is to measure the drift of the thorium lines between a thorium calibration, for which we want a wavelength solution, and a reference thorium calibration, for which a reference wavelength solution already exists. For the work presented in this paper, we derived a reference wavelength solution for the day 2017-11-11, by combining information from a thorium calibration followed by several consecutive Fabry-P\'erot calibrations, which allowed us to obtain a very high-S/N Fabry-P\'erot spectrum, and therefore a precise reference wavelength solution. We then fit on the measured drift of all thorium lines, a smooth two-dimensional polynomial of shape $\Delta X=\sum_{i,j}X^iO^j$, where $i=0,1,2\,\mathrm{and}\,j=0,1,2,3,4$, $X$ corresponds to pixel position in the dispersion direction, and $O$ to order number, to account for variations in the cross-dispersion direction. The corresponding wavelength solution is obtained by shifting the reference wavelength solution using the fitted 15 free parameters two dimensional polynomial. We note that we selected a two dimensional polynomial with $i\le2$ and $j\le4$, as tests using polynomials of smaller or higher dimensions were giving residuals with a larger scatter or no significant improvement, respectively.

%
%

\section{Tailored selection of thorium lines for the new HARPS-N DRS}\label{app:th_line_selection}

With the drift map method used to derive wavelength solutions, described in Sect.~\ref{sect:new_wave_sol} and Appendix~\ref{app:drift_map_method}, we only use 15 free parameters. Thus, we can significantly reduce the number of thorium lines to be used, as long as we are left with thorium lines distributed all over the detector. In addition, the new DRS only fits for a single Gaussian component for each main thorium line, while the old DRS was fitting for several component when blends were known \citep[][]{Lovis-2007b}. If blends vary with time in a different way that the main component, we will observe an asymmetric variation of the main component, which will modify the line's barycenter and therefore induce a spurious RV drift. This spurious change will affect the derived wavelength solution, and therefore the RV stability of the spectrograph on the long-term. We therefore studied the behavior of all the thorium lines used in the drift map method and rejected the ones that were showing strange behaviors.

As described in Sect.~\ref{sect:new_wave_sol}, due to ageing of the thorium-argon lamp used for wavelength solution, the number of lines used over time decreases quite significantly. Starting from the line selection of \citet{Lovis-2007b}, we selected the lines that were used 99\% of the time over the all HARPS-N lifetime\footnote{We note that we wanted to remove all the lines that were not used 100\% of the time, however, a very few bad calibrations were rejecting nearly all the thorium lines. To reject those bad calibrations from our analysis, we therefore lowered slightly the acceptance threshold.}. At the end of this process, we were left with 3061 thorium lines.

\begin{figure}
        \center
	\includegraphics[width=9cm]{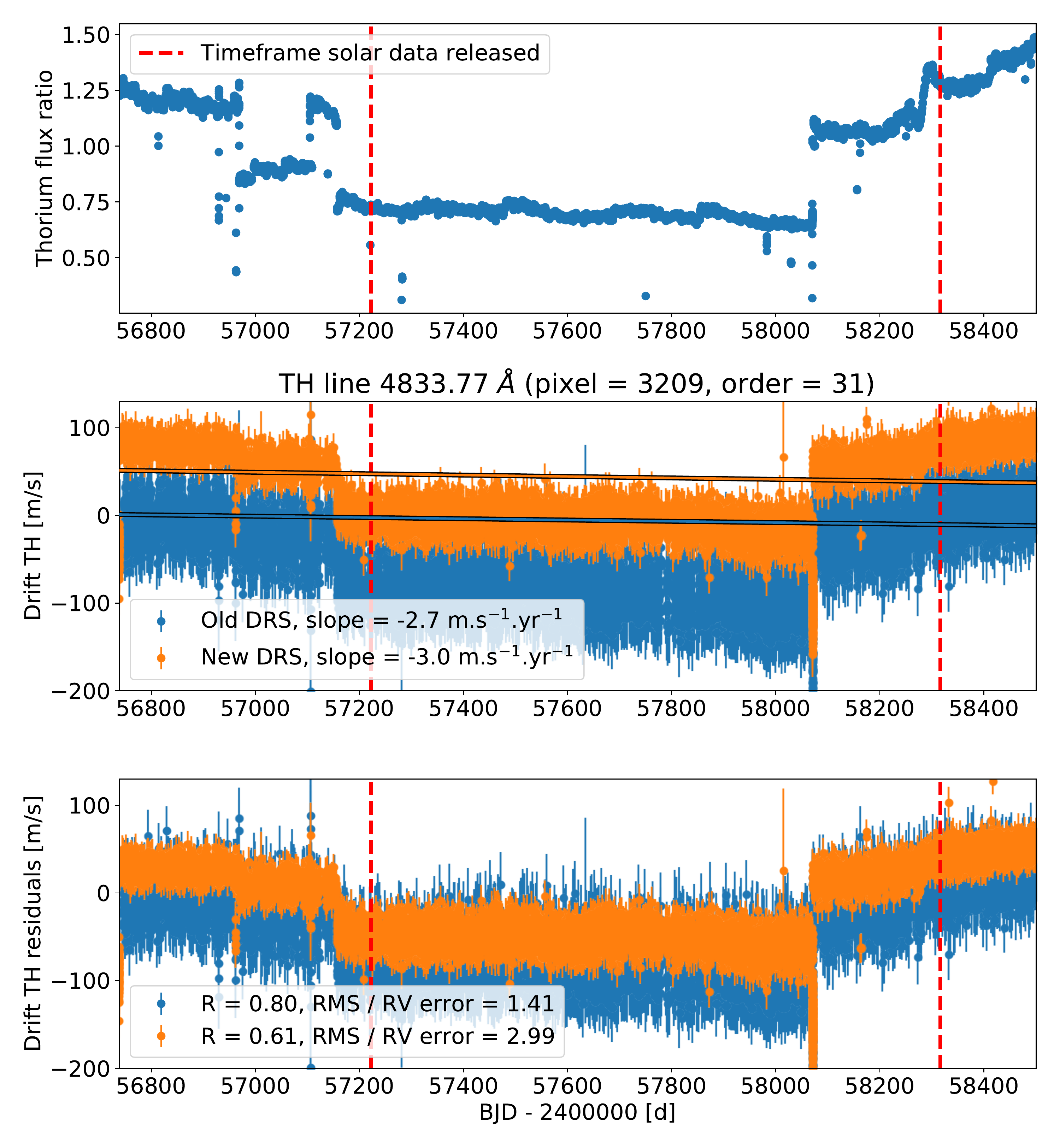}
	\caption{\emph{Top:} Thorium flux ratio as a function of time, which is the total flux emitted by the thorium lamp at a given time relative to a reference time. \emph{Middle:} RV drift of the thorium line at 4833.77 \AA\,measured with the old (blue) and new (orange) DRS. On both data set, we fitted a model corresponding to a linear drift plus a linear correlation with the thorium flux ratio. We reported in the legend the measured slope of the linear drift. \emph{Bottom:} RV drift residuals after subtracting the fitted linear drift. We reported in the legend the R Pearson correlation coefficient between those residuals and the thorium flux ratio, as well as the ratio between the rms of those residuals and the median error in RV for this spectral line, to detect a non Gaussian behavior}.
	\label{fig:thorium_analysis}
\end{figure}

Then we looked at how those remaining lines were behaving as a function of time compared to the median of all the lines. We observed that some lines were showing some drifts over time of dozens of meters-per-second-per-year compared to the median of all the lines, and some others were strongly correlated with the thorium flux ratio, which is the total flux emitted by the thorium lamp at a given time relative to a reference time. The flux ratio is a good proxy to track the ageing of the thorium lamp over time. In the top panel of Fig.~\ref{fig:thorium_analysis}, we show the thorium flux ratio of the thorium lamp used on HARPS-N from the commissioning in 2012 to June 2020. As we can see, the thorium flux ratio changes significantly over time, mainly after BJD=2458070, where we observe a strong linear drift, characteristic of a thorium lamp ageing fast. In the middle and bottom panel of the same figure, we show the RV drift of a thorium line that shows a strong correlation of R=0.80 with thorium flux ratio (in the old DRS). This is likely explained by this thorium line being affected by a blend which vary with the ageing of the lamp. It is clear that such a thorium line shows a strong systematics over time and should not be used.

\begin{figure*}
        \center
    \includegraphics[width=18cm]{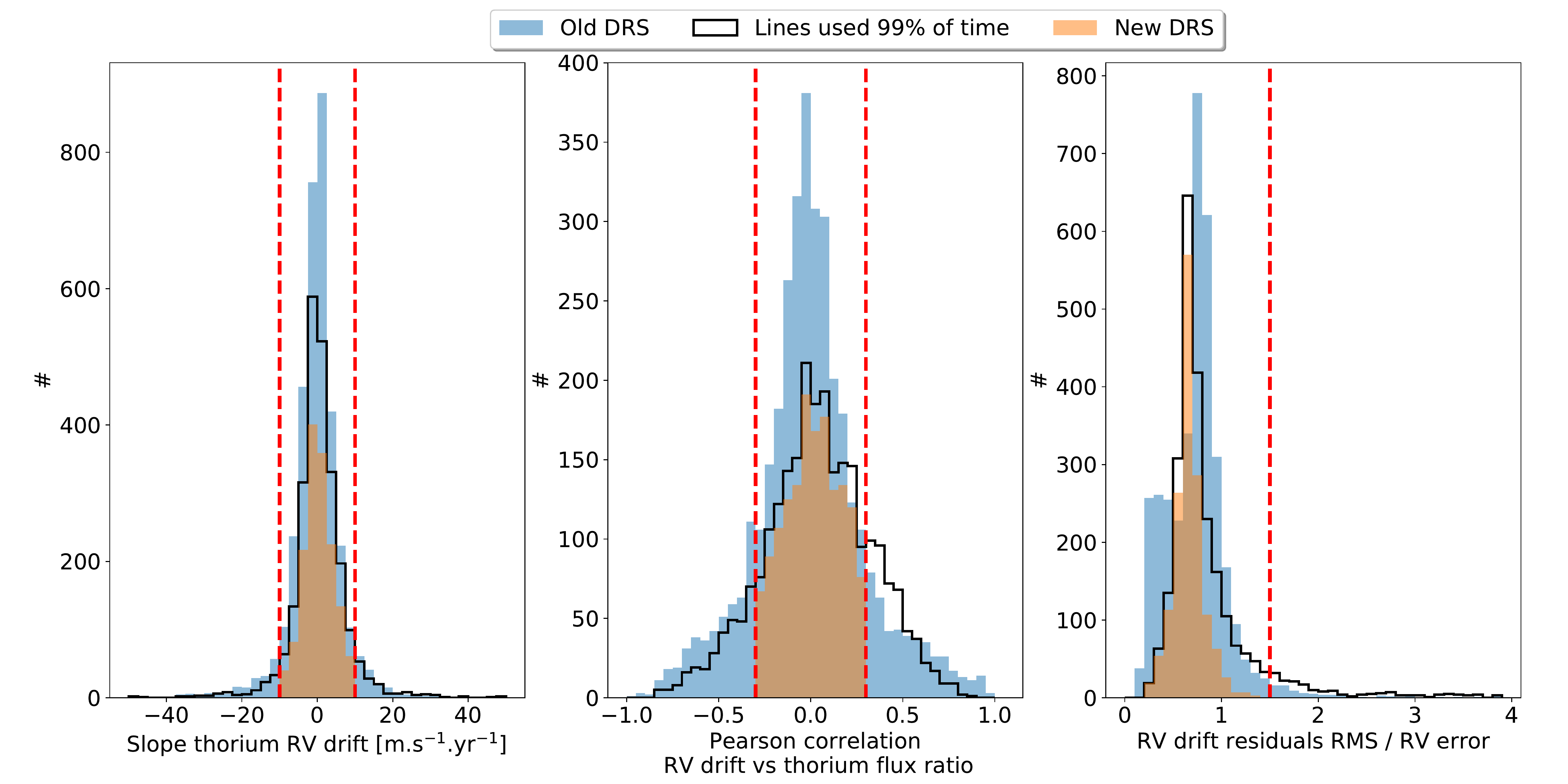}
	\caption{\emph{Left:} Slope measured on the RV drift of all the thorium lines used in the old (blue) and new (orange) DRS. \emph{Middle:} Pearson correlation coefficient between the RV drift of each thorium line and the thorium flux ratio (see Fig.~\ref{fig:thorium_analysis}). \emph{Right:} RMS of the RV drift residuals when subtracting the measured linear drift, divided by the median error in RV. For each subplot, the red vertical dashed lines correspond to the cutoffs used to reject thorium lines showing strong systematics ($\pm$10\ms yr$^{-1}$ for slope, $\pm$0.3 for Pearson correlation coefficient and $<$1.5 for RV residual rms over error in RV). The black histogram corresponds to the selection of lines used 99\% of the time, the orange one to the final selection used in the new DRS, after rejecting lines using the different cutoffs.}
	\label{fig:thorium_selection}
\end{figure*}

To assess the stability of the remaining 3061 thorium lines nearly always used, we measured three different quantities. For each lines, we fitted a model corresponding to a linear drift plus a linear correlation with thorium flux ratio. We then considered the slope of the linear drift and the Pearson correlation coefficient between the RV drift residuals after removing the fitted slope and the thorium flux ratio as two relevant quantities of line stability. We also measured the ratio between the rms of the RV drift residuals and the median RV error, which corresponds to our third quantity. This last quantity probes the presence of remaining signals in the RV drift residuals. We measured those quantities for all the thorium lines used in the old and new DRS, and reported their values in Fig.~\ref{fig:thorium_selection}.

Looking at the different distributions obtained in Fig.~\ref{fig:thorium_selection} for the new DRS thorium line selection, we decided to reject the lines for which the slope was larger than $\pm$10\ms yr$^{-1}$, for which the Pearson correlation coefficient with thorium flux ratio was larger than 0.3, and for which the residual rms over the median error in RV was larger than 1.5. We choose those cutoffs to reject lines with the strongest systematics, while keeping a minimum of nine thorium lines per echelle order to sufficiently constrain the two-dimension polynomial fit performed to derive wavelength solutions (see Appendix~\ref{app:drift_map_method}). After this drastic selection, we were left with 2081 thorium lines.
%

\begin{figure}
        \center
    \includegraphics[width=9cm]{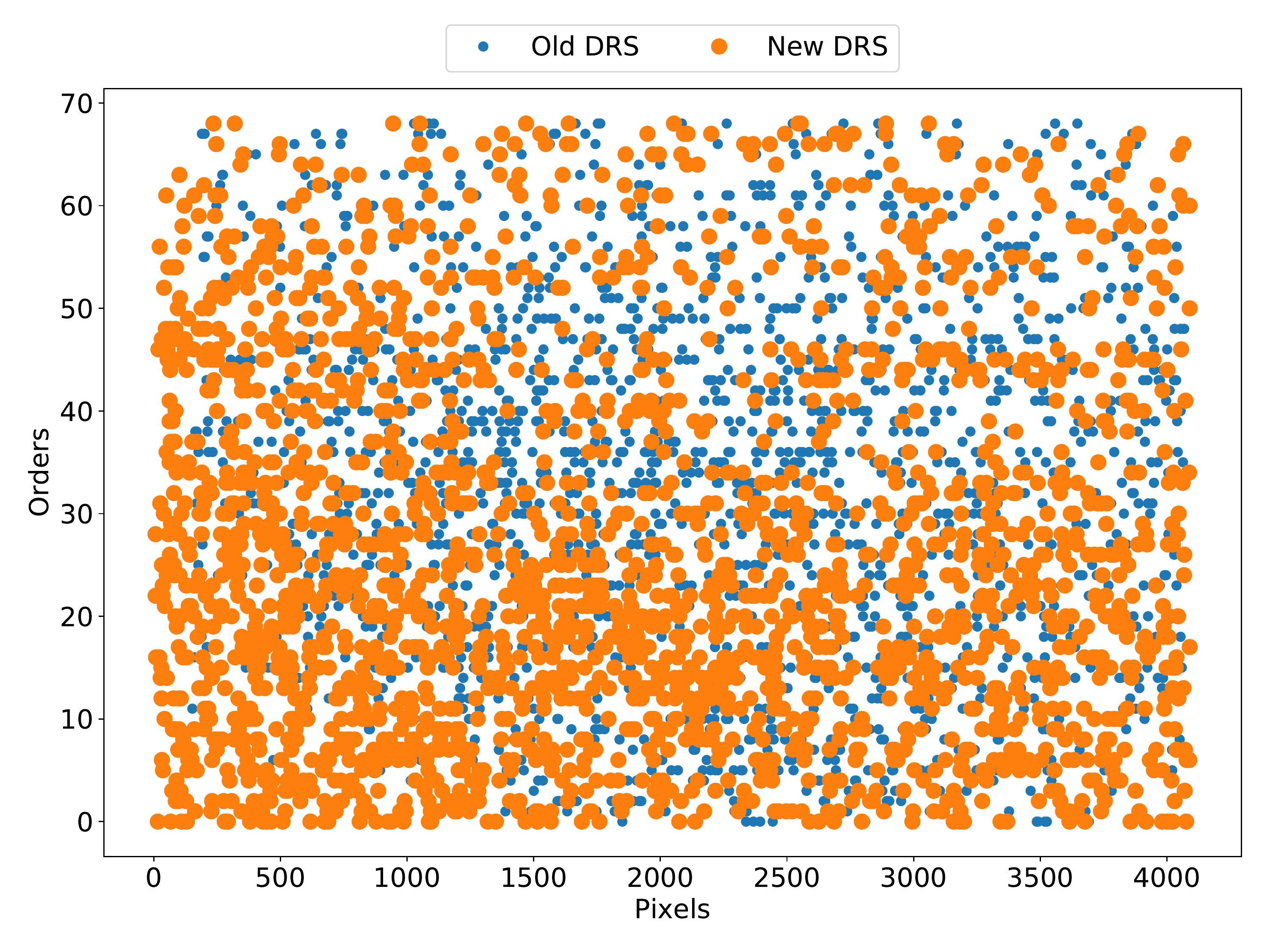}
	\caption{Position of the thorium lines used by the old (blue) and new (orange) DRS as a function of their position on the detector. We note that the position in dispersion direction is in pixel, and the position in cross-dispersion direction in echelle order number.}
	\label{fig:thorium_lines_on_detector}
\end{figure}

Finally, as we used the \citet{Lovis-2007b} catalogue as a starting point, the thorium line wavelengths are coming from \citet{Palmer:1983aa}. \citet{Redman:2014aa} presented a new atlas of thorium lines with more precise wavelength determination and uncertainties. We therefore cross-matched our final selection of lines with this updated catalogue to provide the most up-to-date wavelengths and uncertainties for thorium lines. To do so, we required that a line in our selection had strictly one line in the \citet{Redman:2014aa} atlas within 1\kms. We rejected the 37 lines that were not matching this criteria. Finally, we checked that the ionisation level between the two catalogue was the same. It was not the case for three lines, and therefore we rejected them as well. Our final selection of thorium lines, with updated \citet{Redman:2014aa} wavelengths and uncertainties, is thus composed of 2041 lines that can be found in Table \ref{tab:th_selection}. We can also see the distribution of this final selection of lines over the HARPS-N detector in Fig.~\ref{fig:thorium_lines_on_detector}.
\begin{table}
	\footnotesize
	\caption{Thorium line list used by the new HARPS-N DRS. Wavelengths and their uncertainties are in the vacuum and are coming from \citet{Redman:2014aa}. This table only shows a small fraction of the data that are available in electronic format.}            
	\label{tab:th_selection}    
	\centering                         
	\begin{tabular}{ccccc}       
		\hline\hline
		Wavelength  & Wavelength  & Element & order & pixel \\
		$\quad$[$\AA$]$\quad$ & error [$\AA$] & & & \\ 
		\hline
3875.34184 & 0.00003 & TH1 & 0 & 15.33 \\
3875.96010 & 0.00004 & TH1 & 0 & 65.74 \\
3876.47131 & 0.00007 & TH1 & 0 & 107.50 \\
3876.74453 & 0.00005 & TH1 & 0 & 129.86 \\
3878.56171 & 0.00006 & TH1 & 0 & 279.43 \\
$\cdots$ & $\cdots$ & $\cdots$ & $\cdots$\\
6875.98882 & 0.00014 & TH1 & 68 & 2052.87 \\
6885.03651 & 0.00007 & TH1 & 68 & 2543.73 \\
6885.21375 & 0.00009 & TH1 & 68 & 2553.52 \\
6891.20409 & 0.00009 & TH2 & 68 & 2889.94 \\
6894.15243 & 0.00016 & TH1 & 68 & 3058.97 \\
		\hline
	\end{tabular}
\end{table}
%

\section{Correcting the calcium activity index from ghost contamination}\label{app:rhk_correction}

\begin{figure*}
        \center
        \includegraphics[width=4.28cm]{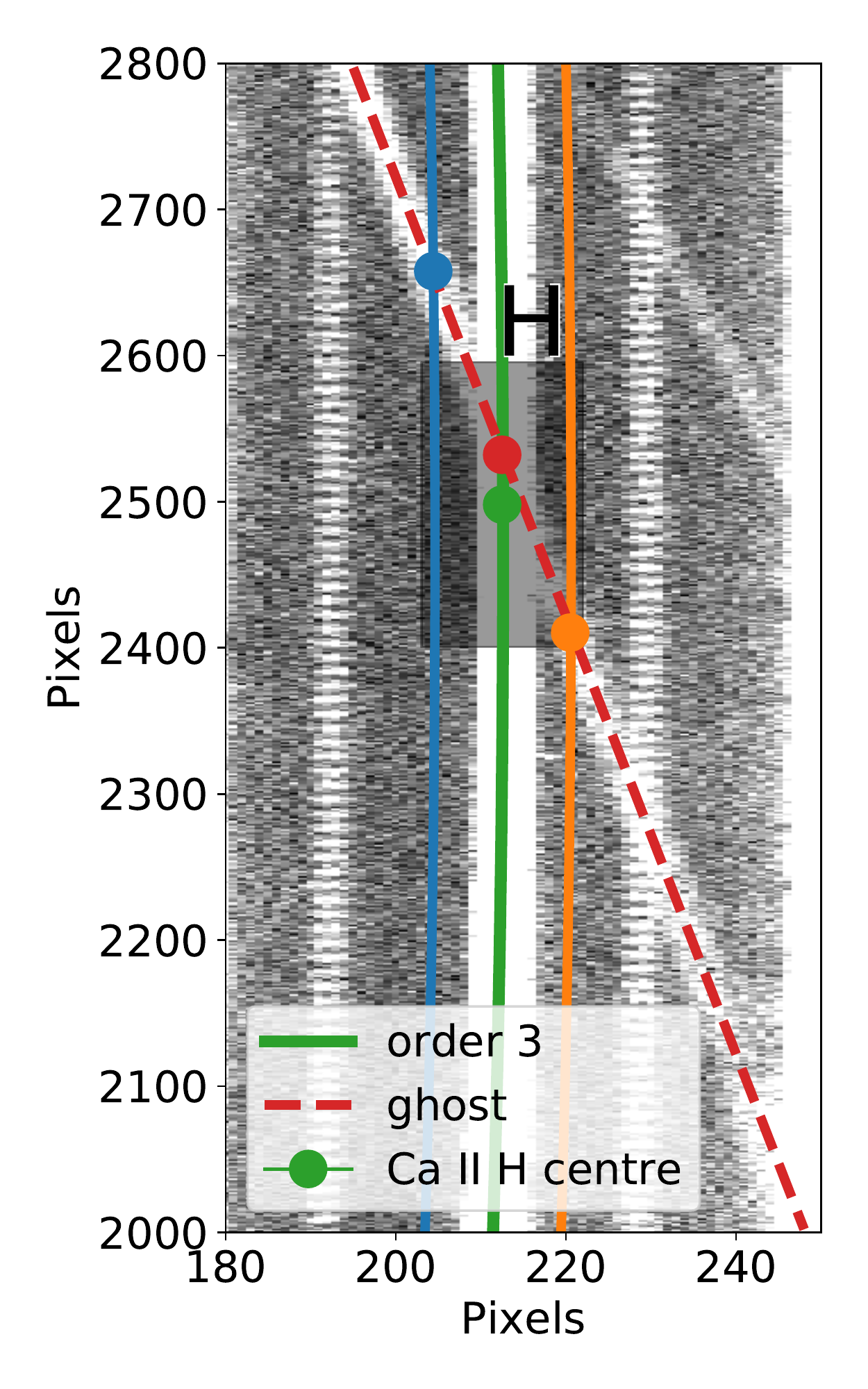}
	\includegraphics[width=13.7cm]{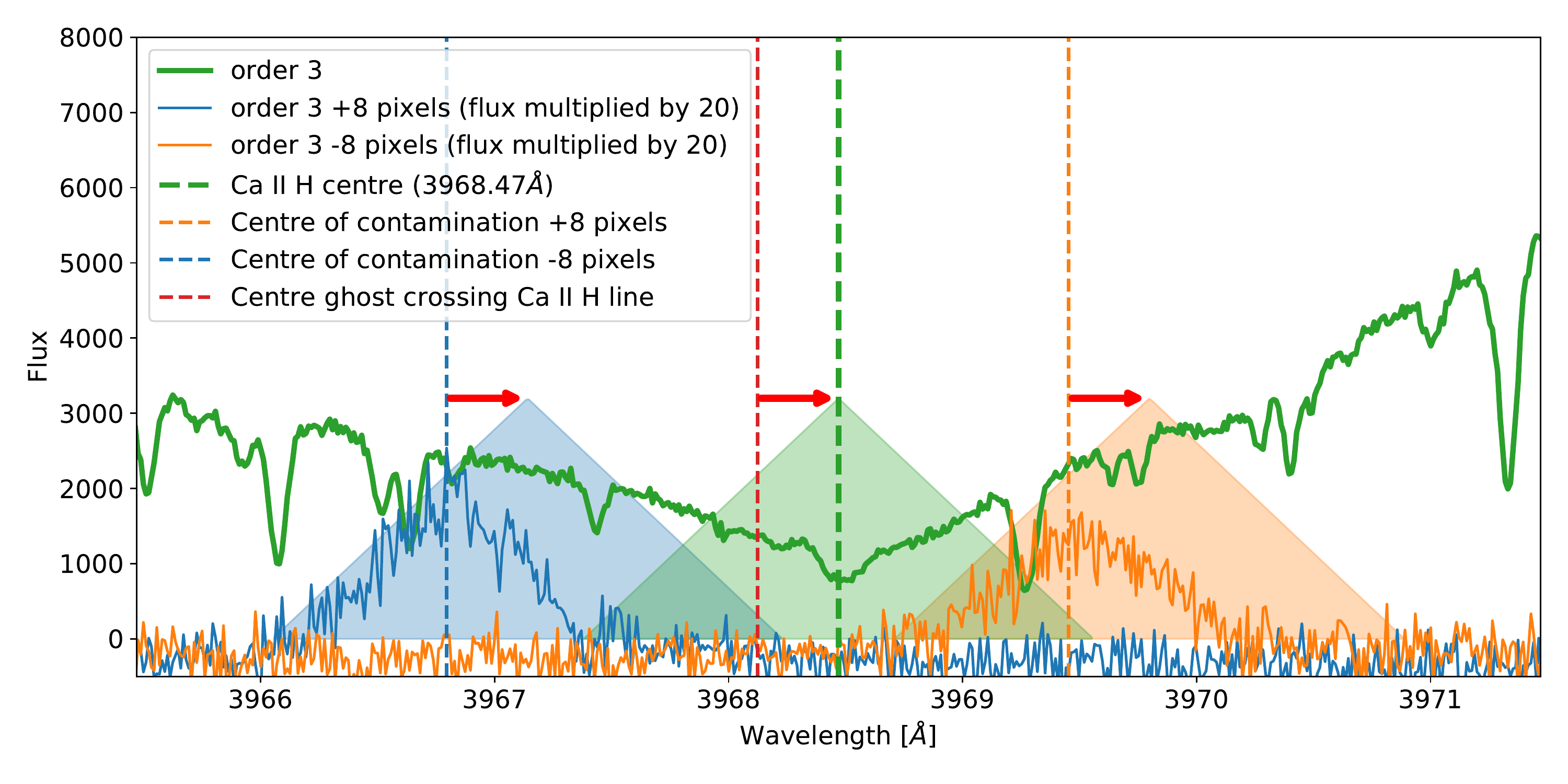}
	\caption{Derivation of the chromospheric emission in the core of the Ca II H line with estimation of the contamination induced by the perturbing ghost. \emph{Left: } Raw image centred on extracted order three with the position of the Ca II H line core. \emph{Right: } The extracted spectra at different positions. The green spectrum is the stellar spectrum extracted from order three (green line on the left panel). The blue and orange spectra are the flux optimally extracted, using the same order profile as order three, minus and plus eight pixels in the cross-dispersion direction on each side of order three (blue and orange lines on the left panel). We note that the extracted fluxes for those spectra have been multiplied by 20 to highlight the contamination from the ghost (red dashed line on the left panel). The vertical green dashed line corresponds to the centre of the Ca II H line core (green dot on the left panel), and the blue and orange vertical dashed lines to the centre of the ghost on both sides of order three (blue and orange dots on the left panel). Assuming that the ghost does not show any curvature, which is a good approximation from one side of the order to the other, we measure the crossing of the ghost and order three (red dot on left panel, and red vertical dashed line on the right panel) as the middle between the orange and blue dots. To measure the chromospheric emission in the core of the Ca II H, the extracted flux from the green spectrum is multiplied with the triangular green response of width $1.09\,\AA$ \citep[][]{Vaughan-1978}. To measure the effective ghost contamination, we multiply the contaminating flux on the blue and the orange spectra with the same triangular response, taking into account that the ghost does not cross order three at the center of the Ca II H line core (the difference is highlighted by the red arrows). Finally, the contamination inside the core of the Ca II H line is taken as the average of the contaminating flux on both sides of the orders.}
	\label{fig:corr_ghost_contam}
\end{figure*}

To calculate the contamination induced by ghosts on the measured flux in the core of the Ca II H and K lines, we optimally extracted the flux, using the profile of the corresponding echelle orders, on both sides of the orders containing the Ca II H and K lines. To select on which pixel to extract the flux, we used the order localisation, and shifted the localisation by plus or minus eight pixels in the cross dispersion direction. We highlighted the localisation of those pixels as blue and orange lines in the left panel of Fig.~\ref{fig:corr_ghost_contam}. As the separation between the science and calibration fibre is at least sixteen pixels, these eight pixel shift allows to extract the flux from ghosts, without being significantly contaminated by the solar or Fabry-P\'erot spectra. Optimally extracting the flux using the profile of the orders allows us to measure the exact flux that contaminates the science fibre. We describe in detail in the caption of Fig.~\ref{fig:corr_ghost_contam} how the ghost contamination is measured in the core of the Ca II H line. We perform the same measurements for the core of the Ca II K line appearing in order one and two of HARPS-N, and finally correct from this contamination before measuring the corrected S-index, and relative \logrhk.

\end{appendix}

\end{document}